\documentclass[twocolumn,preprintnumbers,amsmath,amssymb]{revtex4}
\usepackage{epsfig}
\usepackage{color}
\usepackage[colorlinks,urlcolor=blue,linkcolor=blue,citecolor=red]{hyperref}
\begin{document}
\setlength{\voffset}{1.0cm}
\title{Beyond integrability: Baryon-baryon backward scattering \\ in the massive Gross-Neveu model}
\author{Michael Thies\footnote{michael.thies@gravity.fau.de}}
\affiliation{Institut f\"ur  Theoretische Physik, Universit\"at Erlangen-N\"urnberg, D-91058, Erlangen, Germany}
\date{\today}

\begin{abstract}
Due to integrability, baryon-baryon scattering in the massless Gross-Neveu model at large $N$ features only forward elastic scattering.
A bare mass term breaks integrability and is therefore expected to induce backward elastic scattering as well as inelastic reactions. We confirm these
expectations by a study of baryon-baryon scattering in the massive Gross-Neveu model near the non-relativistic limit. This restriction enables us
to solve the time-dependent Hartree-Fock equations with controlled approximations, using a combination of analytical methods from an effective field theory 
and the numerical solution of partial differential equations.
\end{abstract}
\maketitle
\section{Introduction}
\label{sect1}
The massive Gross-Neveu (GN) model \cite{1} is the quantum field theory of $N$ flavors of Dirac fermions with a scalar-scalar four-fermion interaction and 
Lagrangian 
\begin{equation}
{\cal L} = \sum_{i=1}^N \bar{\psi}^{(i)} \left( i \gamma^{\mu} \partial_{\mu} - m_0 \right)\psi^{(i)} + \frac{1}{2} g^2 \left( \sum_{i=1}^N \bar{\psi}^{(i)} \psi^{(i)} \right)^2.
\label{A1}
\end{equation}
We will only consider the case of 1+1 dimensions, where $g^2$ is dimensionless and the theory is renormalizable.
The bare mass $m_0$ breaks explicitly the discrete chiral symmetry of the original, massless GN model ($\psi \to \gamma_5 \psi, \bar{\psi}\psi \to - \bar{\psi} \psi$)
and renders the model non-integrable. A number of explicit, analytical results have been obtained in the 't~Hooft limit ($N\to \infty, Ng^2=$const.) of the massive GN model in the past,
using semi-classical methods. Thus, the vacuum \cite{2}, baryons \cite{3,4}, multi-baryon bound states \cite{5}, cold and dense matter \cite{6} and the phase
diagram at finite chemical potential and temperature \cite{2,7,8} are by now well understood. Somewhat surprisingly, explicit results for static properties have turned
out to be equally accessible in the massive and massless GN models, despite the fact that only the massless one is integrable. In particular, scalar mean fields for
baryons are transparent and those for inhomogeneous phases of dense matter are finite gap periodic potentials, irrespective of whether the bare mass $m_0$ 
is included or not. 

The situation changes once we look at dynamical problems. In the case of the massless GN model, it has proven possible to solve time-dependent
scattering problems of multi-fermion bound states and write down general results in closed analytical form \cite{9}. The method used was based on a relativistic
version of the time-dependent Hartree-Fock (TDHF) approximation, supposed to become exact in the large $N$ limit.
Since these results show only elastic forward scattering and factorized transmission amplitudes, there is little doubt that integrability is at work here. By contrast,
as pointed out in \cite{10}, a similar ansatz method does not yield self-consistent mean fields for the massive GN model, despite the fact that individual static baryons
can be correctly described. 

The aim of this paper is to elaborate on the difference between integrable, massless and and non-integrable, massive GN models. 
Static properties apparently give no clue about this difference, at least at large $N$. One way of going beyond static properties is to head towards non-equilibrium thermodynamics,
e.g., by studying the bulk viscosity \cite{11}. In that reference one also finds a pedagogical discussion of the (non-)integrability of GN models in terms of Feynman diagrams for
inelastic processes.  Here we propose to follow another route. We generalize previous baryon-baryon scattering calculations of the massless GN model to the massive one,
looking directly for non-forward elastic scattering and inelastic reactions. Clearly, we cannot hope to carry out such studies without numerical computations.
Due to the Dirac sea, it would be very challenging to do a numerical TDHF calculation from scratch. Therefore, in this exploratory study, we set
ourselves a more modest goal. We try to identify the leading order contribution to backward and inelastic scattering in the vicinity of the non-relativistic limit only.
This will enable us to build on a previously developed effective low energy theory for the GN model \cite{12}, while keeping numerical computations manageable.
At the same time, by combining analytical and numerical tools, we hope to get more insight than with a purely numerical approach.

We finish this introduction with a reminder about regularization and renormalization of the massive GN model \cite{2,3}. The Lagrangian (\ref{A1}) has two bare parameters,
$g^2$ and $m_0$. After the regularization/renormalization procedure at large $N$, all observables can be expressed through two physical parameters, $m$ and $\gamma$. 
The relation between the bare quantities, an UV cut-off $\Lambda$ and the physical parameters is given by the vacuum gap equation
\begin{equation}
\frac{\pi}{Ng^2} = \gamma + \ln \frac{\Lambda}{m}, \quad \gamma:= \frac{\pi}{Ng^2} \frac{m_0}{m}.
\label{A2}
\end{equation} 
In the massless GN model ($m_0=0$), the dimensionless coupling constant $g^2$ gets traded for the dynamical fermion mass $m$, an 
example of dimensional transmutation. In the massive model, in addition, the bare mass $m_0$ gets replaced by a physical parameter $\gamma$.
In condensed matter physics where the massless and massive Gross-Neveu models can be used for instance to model trans- and cis-polyacetylene 
\cite{13,14}, $\gamma$ is called ``confinement parameter". It can also be related to the ratio of the dynamical fermion mass at $m_0$ and $m_0=0$,
\begin{equation}
\frac{m[m_0]}{m[0]} = e^{\gamma}
\label{A3}.
\end{equation}
The physical parameters ($m,\gamma$) are renormalization group invariant.

The plan of the present paper is as follows. In Sect.~\ref{sect2}, we derive the TDHF equations for the massive GN model in the vicinity of the 
non-relativistic limit. After introducing an appropriate expansion parameter, in Sect.~\ref{sect3} we simplify these equations further, exhibiting the leading order (LO) 
and next-to-leading order (NLO) equations in detail. Sect.~\ref{sect4} explains our method of solving these equations, whereas numerical results will be
presented in Sect.~\ref{sect5}. We close this paper with a summary and conclusions, Sect.~\ref{sect6}.

\section{TDHF near the non-relativistic limit}
\label{sect2}
Our starting point is the Dirac equation with scalar potential $S$ in 1+1 dimensions, 
\begin{equation}
\left(i \gamma^{\mu} \partial_{\mu}-S \right) \psi = 0.
\label{B1}
\end{equation}
In TDHF theory appropriate for the large $N$ limit of the GN model, $S$ is the self-consistent mean field,
\begin{equation}
S = -g^2 \langle \bar{\psi} \psi \rangle = - g^2 \sum_{\alpha}^{\rm occ} \bar{\psi}_{\alpha} \psi_{\alpha},
\label{B2}
\end{equation}
where the sum runs over all occupied states, i.e., the filled Dirac sea and positive energy valence states.
The aim of the present section is to set up the first two terms of a systematic, non-relativistic approximation to the full TDHF problem.

Using the Dirac-Pauli representation ($\gamma^0 = \sigma_3, \gamma^1 = i \sigma_2, \gamma_5 = \sigma_1$) and pulling out the fast factor $e^{-imt}$ from the spinor $\psi$,
we cast Eq.~(\ref{B1}) into the Hamiltonian form
\begin{equation}
i \partial_t  \left( \begin{array}{c}  \psi_1 \\ \psi_2 \end{array} \right) = \left( \begin{array}{cc} S-m & -i \partial_x \\ -i \partial_x & -S-m \end{array} \right) 
 \left( \begin{array}{c}  \psi_1 \\ \psi_2 \end{array} \right).
\label{B3}
\end{equation}
Next, we eliminate the ``small" component $\psi_2$ formally from (\ref{B3}),
\begin{equation}
i \partial_t \psi_1 = (S-m) \psi_1 - \partial_x \frac{1}{S+m+i\partial_t} \partial_x \psi_1,
\label{B4}
\end{equation}
use the non-relativistic expansion
\begin{equation}
\frac{1}{S+m+i \partial_t} \approx \frac{1}{2m} - \frac{(S-m+i\partial_t)}{4m^2}
\label{B5}
\end{equation}
and arrive at a Schr\"odinger-type equation for the ``large" component $\psi_1$,
\begin{equation}
i \partial_t \psi_1 = \left( S-m- \frac{\partial_x^2}{2m}\right) \psi_1 + \partial_x \frac{\left(S-m+i\partial_t \right)}{4m^2} \partial_x \psi_1.
\label{B6}
\end{equation}
Replacing $i\partial_t \psi_1$ on the right hand side by the LO expression $(S-m- \partial_x^2/2m)\psi_1$, we find
\begin{eqnarray}
i \partial_t \psi_1  & = &   \left( S-m- \frac{\partial_x^2}{2m} \right) \psi_1 - \frac{\partial_x^4}{8m^3} \psi_1 
\nonumber \\
& + & \left[ \partial_x \left( \frac{S-m}{4m^2} \right) \partial_x  + \partial_x^2 \left( \frac{S-m}{4m^2} \right) \right] \psi_1.
\label{B7}
\end{eqnarray}
The first term on the right is of LO, the second term the NLO correction to the relativistic kinetic energy,
\begin{equation}
\sqrt{m^2+p^2} -m - \frac{p^2}{2m} \approx  - \frac{p^4}{8m^3}.
\label{B8}
\end{equation}
The third term is non-hermitean, reflecting the fact that after elimination of the lower component $\psi_2$, the norm of the upper component $\psi_1$ is not conserved. To 
NLO, the conserved charge is
\begin{equation}
Q = \int dx \left( |\psi_1|^2 + | \psi_2|^2 \right) \approx \int dx \psi_1^* \left( 1- \frac{\partial_x^2}{4m^2} \right) \psi_1.
\label{B9}
\end{equation}
Accordingly, we redefine the Hamiltonian and the wave functions as follows,
\begin{eqnarray}
H  \longrightarrow  \tilde{H} &  = &  \left( 1- \frac{\partial_x^2}{4m^2} \right)^{1/2} H  \left( 1- \frac{\partial_x^2}{4m^2} \right)^{-1/2}
\nonumber \\ 
& \approx &  H - \frac{1}{8m^2} \left[ \partial_x^2, H \right],
\label{B10} \\
\psi_1 \longrightarrow  \tilde{\psi}_1 & = &  \left( 1- \frac{\partial_x^2}{4m^2} \right)^{1/2} \psi_1 \approx \left( 1- \frac{\partial_x^2}{8m^2} \right) \psi_1.
\nonumber
\end{eqnarray}
This yields the amended version of the Schr\"odinger equation (\ref{B7}), now with a manifestly hermitean Hamiltonian,
\begin{eqnarray}
i \partial_t \tilde{\psi}_1 & = & \left( - \frac{\partial_x^2}{2m} + S - m - \frac{\partial_x^4}{8m^3} \right.
\nonumber \\
&  + & \left.  \frac{1}{8 m^2} \left\{ \partial_x, \left\{ \partial_x, S-m \right\}\right\}
\right) \tilde{\psi}_1.
\label{B11}
\end{eqnarray}
We have generated the analogue of the Darwin term for a scalar potential (for a vector potential, replace the anti-commutators by commutators.) 
The reader will have noticed that this calculation follows closely the textbook evaluation of the fine structure of the hydrogen atom, see e.g. \cite{15}, 
except for the missing spin-orbit term in 1+1 dimensions. 

What we have done so far is the non-relativistic reduction of the Dirac equation with a classical potential, including fine structure
corrections. From the hydrogen atom, we already know that this is not the whole story: the Lamb shift is still missing. It arises from vertex corrections and
vacuum polarization due to Dirac sea. Fortunately, we have the tools at our hands to include such quantum-field corrections systematically as well.
To this end we turn to an ``effective no-sea theory" of the massive GN model, derived in Ref.~\cite{12} by ``integrating out" all negative energy states. 
Referring to \cite{12} for the technical details and derivations, we immediately jump to the results. 
Keeping only the first two terms of a systematic expansion, one finds for the mean field of the massive GN model  
\begin{eqnarray}
S - m & = & \sigma - \frac{1}{12 m^2} \frac{1}{(1+\gamma)} \partial_{\mu}\partial^{\mu} \sigma - \frac{1}{2m} \frac{1}{(1+\gamma)} \sigma^2,
\nonumber \\
\sigma & = & -  \frac{\pi}{(1+\gamma)} \sum_{\ell} \nu_{\ell} \bar{\psi}_{\ell} \psi_{\ell}.
\label{B12}
\end{eqnarray}
Let us briefly explain the various symbols and terms. To LO, $S=m+\sigma$. Here, $m$ is the dynamical fermion mass which would arises from the 
Dirac sea in the full HF calculation but has to be put in by hand in the effective theory. The scalar field $\sigma$ --- the field of the $\sigma$ meson --- 
has a similar self-consistent structure as $S$ in (\ref{B2}). But here, the sum over occupied states only extends over the positive energy valence levels.
The coefficient $\nu_{\ell}=N_{\ell}/N$ denotes the occupation fraction of the $\ell$-th bound state, a continuous parameter in the limit $N \to \infty$.
The bare coupling constant $g^2$ gets replaced by a renormalized coupling constant $g_{\rm eff}^2 = - \frac{\pi}{N(1+\gamma)}$.
This value can be derived from the tadpole contribution to the fermion self energy by projecting out the negative energy states, see Ref.~\cite{12}.
The other two terms on the right hand side are NLO and arise from vacuum polarization effects. As explained in Ref.~ \cite{12}, one needs
to re-sum an infinite number of diagrams contributing to the $\sigma$-meson propagator. In this process, the bare coupling
constant drops out, so that the result is truly non-perturbative in $g^2$. To arrive at the final form, one expands the inverse $\sigma$-propagator in powers of 
$k^2$, so that the truncation underlying Eq.~(\ref{B12}) is indeed consistent with the non-relativistic reduction of the Dirac equation.

The task is now to solve Eq.~(\ref{B11}) for all bound states, using as self-consistency condition, Eq.~(\ref{B12}). Although these equations look
more complicated than the original TDHF equations (\ref{B1},\ref{B2}), they are much easier to solve. It is sufficient to determine
positive energy bound states self-consistently, rather than the whole Dirac sea plus valence levels. Thus, the structure resembles that of a
non-relativistic TDHF problem, albeit with a more complicated interaction and self-consistency condition. Relativistic corrections are included by means 
of the fine structure terms as well as the higher order corrections to the no-sea effective theory. There are no more divergences, as regularization and
renormalization have already been performed when deriving (\ref{B12}). Since the scalar condensate $\bar{\psi}\psi$ in (\ref{B12}) still contains the original
two-component spinors $\psi$ rather than $\tilde{\psi}_1$, the equations are not yet in a form ready to be solved. Further simplifications will appear once 
we introduce a small expansion parameter and truncate all the equations consistently, the goal of the following section.
 
\section{Expansion parameter, LO and NLO TDHF equations}
\label{sect3}

In order to arrive at a tractable set of equations and to avoid the unnecessary computation of complicated higher order terms, we introduce a formal
expansion parameter $\epsilon$. The regime we are interested in is characterized by $v \sim \epsilon, y \sim \epsilon$, with $v$ the baryon
velocity and $y$ the (inverse) baryon size parameter \cite{3,4}. The first condition is self-evident for a non-relativistic expansion. The 2nd one induces a matching 
non-relativistic expansion for the internal structure of the baryon.
The characteristic exponential in the single baryon is exp$[2y(x-vt)]$, hence we treat $\partial_x \sim \epsilon,
\partial_t \sim \epsilon^2$. Guided by the single baryon results, we assume the following expansions for $S-m$ and the spinors,
\begin{eqnarray}
S-m & = & \epsilon^2 S^{(2)} + \epsilon^4 S^{(4)},
\nonumber \\
\tilde{\psi}_1 & = & \sqrt{\epsilon} \left( \tilde{\psi}_1^{(0)} + \epsilon^2 \tilde{\psi}_1^{(2)} \right),
\nonumber \\
\tilde{\psi}_2 & = & \epsilon^{3/2} \left( \tilde{\psi}_2^{(0)}+ \epsilon^2 \tilde{\psi}_2^{(2)} \right).
\label{C1}
\end{eqnarray}
The small component $\tilde{\psi}_2$ will be needed later on for the condensate. 
Inserting these expressions into the Dirac equation (\ref{B11}) with $S-m$ from (\ref{B12}) and equating powers of $\epsilon$ then yields the following
LO and NLO equations,
\begin{eqnarray}
i \partial_t \tilde{\psi}_{k,1}^{(0)} & = & \left( - \frac{\partial_x^2}{2m} + S^{(2)} \right) \tilde{\psi}_{k,1}^{(0)} \ \ ({\rm LO})
\nonumber \\
i \partial_t \tilde{\psi}_{k,1}^{(2)} & = & \left( - \frac{\partial_x^2}{2m} + S^{(2)} \right) \tilde{\psi}_{k,1}^{(2)} 
 + \left( S^{(4)} - \frac{\partial_x^4}{8m^3} \right.
\nonumber \\
& + & \left. \frac{1}{8m^2} \left\{ \partial_x, \left\{ \partial_x, S^{(2)} \right\} \right\} \right) \tilde{\psi}_{k,1}^{(0)} \ \  ({\rm NLO})
\label{C2}
\end{eqnarray}
We have added a subscript $k$ labeling the bound states, as required by the general TDHF problem.

We still need the relation between $S^{(2)}, S^{(4)}$ and the spinors $\tilde{\psi}_{k,1}^{(0)}, \tilde{\psi}_{k,1}^{(2)}$ arising from the self-consistency 
condition (\ref{B12}) and expansions (\ref{C1}). For each baryon, there is a size parameter $y_k$ and an occupation fraction $\nu_k$ related in a
non-linear fashion, namely \cite{3,4}
\begin{equation}
\frac{\nu_k}{2} = \frac{\arcsin y_k}{\pi} + \frac{\gamma}{\pi} \frac{y_k}{\sqrt{1-y_k^2}}.
\label{C3}
\end{equation}
Since $y_k \sim \epsilon$, $\nu_k$ can be expanded as
\begin{equation}
\nu_k = \epsilon \nu_k^{(1)} + \epsilon^3 \nu_k^{(3)}
\label{C4}
\end{equation}
with
\begin{eqnarray}
\nu_k^{(1)} & = & \frac{2 y_k (1+\gamma)}{\pi},
\nonumber \\
\nu_k^{(3)} & = & \frac{y_k^3(1+3\gamma)}{3\pi}.
\label{C5}
\end{eqnarray}
Inserting Eqs.~(\ref{C1},\ref{C4},\ref{C5}) into Eq.~(\ref{B12}) yields
\begin{eqnarray}
\sigma & = & \epsilon^2 \sigma^{(2)} + \epsilon^4 \sigma^{(4)},
\nonumber \\
\sigma^{(2)} & = & - \frac{\pi}{1+\gamma} \sum_{\ell}  \nu_{\ell}^{(1)} | \tilde{\psi}_{\ell,1}^{(0)} |^2 ,
\label{C6} \\
\sigma^{(4)} & = & -  \frac{\pi}{1+\gamma} \sum_{\ell} \left(  \nu_{\ell}^{(3)} | \tilde{\psi}_{\ell,1}^{(0)} |^2 - \nu_{\ell}^{(1)} | \tilde{\psi}_{\ell,2}^{(0)} |^2\right)
\nonumber \\
& - &  \frac{\pi}{1+\gamma} \sum_{\ell} \nu_{\ell}^{(1)} \left( \tilde{\psi}_{\ell,1}^{(0)} \tilde{\psi}_{\ell,1}^{(2)*} 
 +  \frac{1}{8m^2} \tilde{\psi}_{\ell,1}^{(0)} \partial_x^2 \tilde{\psi}_{\ell,1}^{(0)*}
+ {\rm c.c.} \right) 
\nonumber
\end{eqnarray}
where 
\begin{equation}
\tilde{\psi}_{\ell,2}^{(0)} = - \frac{i}{2m} \partial_x \tilde{\psi}_{\ell,1}^{(0)}.
\label{C7}
\end{equation}
The various terms in (\ref{C6}) can be understood as follows. The scalar density in our representation of the $\gamma$-matrices is
\begin{equation}
\bar{\psi} \psi = |\psi_1|^2 - |\psi_2|^2.
\label{C8}
\end{equation}
Remembering the different powers of $\epsilon$ in (\ref{C1}), this explains the LO term $\sigma^{(2)}$. To this order, the difference between $\psi$ and
$\tilde{\psi}$ does not matter. $\sigma^{(4)}$ contains all NLO terms, taking into account the fact that $\nu_k$ has the expansion (\ref{C4}).
The last line in Eq.~(\ref{C6}) is due to interference terms between $\tilde{\psi}_{\ell,1}^{(0)}$ and $\tilde{\psi}_{\ell,1}^{(2)}$, the only NLO contribution to
$|\tilde{\psi}_{\ell,1}|^2$, and a derivative term coming from the transformation from $\psi$ to $\tilde{\psi}$, see (\ref{B10}). In the $|\psi_2|^2$ term in (\ref{C8}), only the lowest
order is needed, so that we get away with expression (\ref{C7}).

According to Eqs.~(\ref{B12},\ref{C1}) and (\ref{C6}), the relationship between $S^{(2,4)}$ and $\sigma^{(2,4)}$ is
\begin{eqnarray}
S^{(2)} & = & \sigma^{(2)},
\label{C9} \\
S^{(4)} & = & \sigma^{(4)} + \frac{1}{12m^2} \frac{1}{(1+\gamma)} \partial_x^2 \sigma^{(2)} - \frac{1}{2m} \frac{1}{(1+\gamma)} (\sigma^{(2)})^2.
\nonumber
\end{eqnarray}
Time derivatives in $\partial_{\mu}\partial^{\mu} \sigma$ are of higher order than what is needed here. Eqs.~(\ref{C2},\ref{C6},\ref{C9}) together with the
normalization conditions are a closed set of equations determining the positive energy bound state spinors in LO and NLO.

Before going on, it is useful to ease the notation.
Since we can now express everything through the ``large" components $\psi_1$, we drop the subscript 1 from all the spinors. We also omit the 
tilde on all wave functions, replace $\sigma^{(2)}$ everywhere by $S^{(2)}$ and insert $\nu_k^{(1,3)}$. The basic TDHF equations then read to LO
\begin{widetext}
\begin{equation}
i \partial_t \psi_{k}^{(0)}   =    \left( - \frac{\partial_x^2}{2m} + S^{(2)} \right) \psi_{k}^{(0)},
\quad S^{(2)}   =    -  \sum_{\ell} 2 y_{\ell}  | \psi_{\ell}^{(0)} |^2  \quad ({\rm LO}),
\label{C10}
\end{equation}
and to NLO
\begin{eqnarray}
i \partial_t \psi_{k}^{(2)} & = & \left( - \frac{\partial_x^2}{2m} + S^{(2)} \right) \psi_{k}^{(2)}
 + \left( S^{(4)} - \frac{\partial_x^4}{8m^3} 
 +   \frac{1}{8m^2} \left\{ \partial_x, \left\{ \partial_x, S^{(2)} \right\} \right\} \right) \psi_{k}^{(0)} \quad ({\rm NLO}),
\nonumber \\
S^{(4)} &  = &   \sigma^{(4)} + \frac{1}{12m^2} \frac{1}{(1+\gamma)} \partial_x^2 S^{(2)}
 -   \frac{1}{2m} \frac{1}{(1+\gamma)} (S^{(2)})^2,
\nonumber \\
\sigma^{(4)} & = & -   \sum_{\ell} \left(  \frac{y_{\ell}^3}{3} \frac{1+3 \gamma}{1+\gamma} | \psi_{\ell}^{(0)} |^2 -   \frac{y_{\ell}}{2m^2} | \partial_x \psi_{\ell}^{(0)} |^2\right)
  -   \sum_{\ell} 2 y_{\ell}  \left( \psi_{\ell}^{(0)} \psi_{\ell}^{(2)*} + \frac{1}{8m^2} \psi_{\ell}^{(0)} \partial_x^2 \psi_{\ell}^{(0)*}
+ {\rm c.c.} \right)
\label{C11}
\end{eqnarray}
\end{widetext}
These equations are valid for arbitrary $\gamma$. 

Let us first look at the LO equation (\ref{C10}) which has the form of the multi-component non-linear Schr\"odinger (NLS) equation \cite{17}. 
We notice that it does not contain $\gamma$ at all, so that the same equation is valid independently of the bare fermion mass.
How is this possible? The answer is the same as for static baryons: The wave functions and mean fields for a baryon in the massive theory are identical to those for a baryon
in the massless theory, but for a different fermion number. Indeed, $\gamma$ still appears in Eq.~(\ref{C3}) relating the occupation fraction and the size parameter. In the potential $S^{(2)}$, this
is cancelled exactly by the $\gamma$-dependence of the effective coupling constant.
Thus the TDHF equation of the massive GN model reduces to coupled NLS equations, independently of the bare mass.
In this strict non-relativistic limit, we can solve baryon-baryon scattering problems by simply taking the solution of the massless model 
and expanding it in powers of $y,v$ to LO. At this level, there will be neither backward scattering nor any inelastic reaction. The model does not yet loose integrability.

We now turn to NLO, Eq.~(\ref{C11}). The potential $S^{(4)}$ consists of several terms with different $\gamma$-dependencies. Somewhat surprisingly, one can eliminate the $\gamma$ dependence
as follows. Let us denote the solutions of Eq.~(\ref{C11}) at $\gamma=0$ (chiral limit) by hatted quantities $\hat{\psi}_k^{(2)}, \hat{S}^{(4)}$ (we just saw that $\hat{\psi}_k^{(0)} = \psi_k^{(0)}, \hat{S}^{(2)} =S^{(2)}$).
We then make the following ``scaling" ansatz for the NLO quantities,
\begin{eqnarray}
\psi_k^{(2)} & = & \hat{\psi}_k^{(2)} + \frac{\gamma}{1+\gamma}  \chi_k^{(2)},
\nonumber \\
S^{(4)} & = & \hat{S}^{(4)} + \frac{\gamma}{1+\gamma} s^{(4)}.
\label{C12}
\end{eqnarray}
Inserting (\ref{C12}) into (\ref{C11}) and using the fact that the hatted spinors satisfy Eq.~(\ref{C11}) at $\gamma=0$, we arrive at our final set of equations
\begin{eqnarray}
i \partial_t \chi_k^{(2)} & = & \left( - \frac{\partial_x^2}{2m} + S^{(2)} \right) \chi_k^{(2)}  + s^{(4)} \psi_k^{(0)},
\nonumber \\
s^{(4)} & = & -  \sum_{\ell} 2 y_{\ell} \left( \psi_{\ell}^{(0)} \chi_{\ell}^{(2)*} + \psi_{\ell}^{(0)*} \chi_{\ell}^{(2)} \right)  + {\cal F},
\label{C13} \\
{\cal F} &  = &  - \frac{2}{3} \sum_{\ell} y_{\ell}^3 |\psi_{\ell}^{(0)} |^2 - \frac{1}{12m^2} \partial_x^2 S^{(2)} + \frac{1}{2m} \left( S^{(2)} \right)^2
\nonumber
\end{eqnarray}
now independent of $\gamma$. They have the structure of a system of inhomogeneous, linear PDE's. 
The only input needed from the massless GN model are the LO quantities $\psi_k^{(0)}, S^{(2)}$. 
Since the solutions of Eqs.~(\ref{C11}) in the chiral limit can be inferred from the analytically known exact solutions simply by Taylor expansion, there is no need to ever
solve the complicated equations (\ref{C11}) at $\gamma=0$. Solving Eq.~(\ref{C13}) yields the solutions for all values of $\gamma$ at once. 

The inhomogeneity has now been isolated in the term ${\cal F}\psi_k^{(0)}$. At this point, before turning to the solution, we can perform a non-trivial
consistency check of our formalism. Since we know that all corrections must vanish in the static case, we expect that ${\cal F}=0$ if all solitons are at rest.
In the static case, the $x$ and $t$ variables can be separated by going to stationary states,
\begin{equation}
\psi_k^{(0)}(x,t) = e^{i \epsilon_k t} \phi_k^{(0)}(x),
\label{C14}
\end{equation}
where $\phi_k^{(0)}(x)$ is the eigenfunction of the LO Hamiltonian with eigenvalue $\epsilon_k$,
\begin{equation}
\left( - \frac{1}{2m} \partial_x^2 + S^{(2)} \right) \phi_k^{(0)}(x) = \epsilon_k \phi_k^{(0)}(x), \quad \epsilon_k = - \frac{m y_k^2}{2}.
\label{C15}
\end{equation}
$S^{(2)}$ is defined as above, Eq.~(\ref{C10}). Choosing the $\phi_k^{(0)}$ to be real, it becomes
\begin{equation}
S^{(2)} = - \sum_{\ell} 2y_{\ell} (\phi_k^{(0)})^2.
\label{C16}
\end{equation}
A simple exercise with Computer Algebra is the following: Evaluate $\partial_x^2 S^{(2)}$ and eliminate the 2nd derivatives of the time independent
wave functions with the help of the NLS equation (\ref{C15},\ref{C16}). Then differentiate the resulting ${\cal F}$ with respect to $x$. If one
eliminates once again $\partial_x^2 \phi_k^{(0)}$ using (\ref{C15}), one finds that $\partial_x {\cal F}=0$, i.e., ${\cal F}$ is
constant in the static case. Since it obviously vanishes asymptotically, it has to be identically zero. For time dependent problems
this proof fails and we cannot avoid solving the inhomogeneous, linear system of PDE's, Eq.~(\ref{C13}).

\section{Method of solution}
\label{sect4}

So far, things are valid for any number and type of scatterers. Let us first look at the trivial case of a single baryon, where we
would expect no correction at all, since there is a frame in which it is static. In this case the inhomogeneous term in Eq.~(\ref{C13}) must vanish, so that the system
of equations admits the trivial solution $\chi_{1}^{(2)}=0$. As the single baryon of the massive GN model has the same structure as the original Dashen-Hasslacher-Neveu (DHN)
baryon of the massless model \cite{16}, we shall refer to the single baryon as DHN baryon. 
For the single DHN baryon with size parameter $y$ and velocity $v$ and using units where $m=1$ from now on, we have
\begin{equation}
S^{(2)}  =    -2y | \psi_1^{(0)} |^2 = -4 y^2 \frac{U_1}{(1+U_1)^2}, \ \  U_1  =  e^{2y(x-vt)}
\label{D1}
\end{equation}
This yields indeed ${\cal F}=0$ when inserted into (\ref{C13}).

The first non-trivial example is scattering of two DHN baryons. In view of the exploratory character of our study, we 
simplify things as much as possible. We take two identical DHN baryons ($y_1=y_2=y$) in the center-of-mass (cm) frame ($v_1 = - v_2 = v$). Eq.~(\ref{C13}) then becomes
\begin{widetext}
\begin{eqnarray}
\left( i \partial_t + \frac{1}{2} \partial_x^2- S^{(2)} \right) \chi_k^{(2)} & = &  -2 y \psi_k^{(0)} \sum_{\ell} 
\left( \psi_{\ell}^{(0)} \chi_{\ell}^{(2)*} + \psi_{\ell}^{(0)*} \chi_{\ell}^{(2)} \right) + {\cal F} \psi_k^{(0)},
\nonumber \\
{\cal F} & = &  \frac{y^2}{3}S^{(2)} - \frac{1}{12} \partial_x^2 S^{(2)} + \frac{1}{2} \left( S^{(2)} \right)^2
\label{D2}
\end{eqnarray}
The solution of the LO problem is well known \cite{9,10},
\begin{eqnarray}
S^{(2)} &  = & -4 y^2 v^2 \frac{v^2 U_1(1+4 U_2 + U_1U_2)+(v^2+y^2)U_2(1+U_1U_2)}{{\cal D}^2},
\nonumber \\
\psi_1^{(0)} & = & -i v \sqrt{2y U_1} \frac{v+(v+iy)U_2}{\cal D} e^{i(t(y^2-v^2)/2+vx)} ,
\nonumber \\
\psi_2^{(0)} & = & -i \frac{v-iy}{\sqrt{y^2+v^2}} v \sqrt{2y U_2} \frac{(v+iy)+vU_1}{\cal D} e^{i(t(y^2-v^2)/2-vx)} ,
\nonumber \\
{\cal D} & = & v^2(1+U_1)(1+U_2) + y^2 U_2 ,
\nonumber \\
U_1 & = & \lambda^{-1} e^{2y(x-vt)}, \quad U_2 = \lambda e^{2y(x+vt)}, \quad \lambda= \frac{v}{\sqrt{v^2+y^2}}.
\label{D3}
\end{eqnarray}
\end{widetext}
The prefactors $\lambda^{-1}, \lambda$ in $U_{1,2}$ just shift the variables $x,t$ by a constant and have been chosen so as to slightly simplify the  
NLO terms in the chiral limit. The entries in Eq.~(\ref{D3}) can be inferred from the known exact solution by transforming from the chiral representation of Dirac matrices 
used in Ref.~\cite{9} to the Dirac-Pauli representation, replacing $y\to \epsilon y, v\to \epsilon v, x\to \epsilon^{-1}x, t \to \epsilon^{-2} t$ and performing a Taylor expansion in $\epsilon$
to LO. The phases of the two bound state spinors are of course arbitrary, but must be chosen such that there are no linear terms in $\epsilon$ in Eq.~(\ref{C1}). 
One can easily verify that the quantities in (\ref{D3}) satisfy the LO equations and the normalization condition. Using this input, ${\cal F}$ turns out to be significantly simpler than its three constituent terms,
\begin{equation}
{\cal F}  =  \frac{16 v^4 y^4 U_1U_2}{{\cal D}^2}.
\label{D4}
\end{equation}
This is non-zero only when the two baryons overlap. Asymptotically, ${\cal F}$ vanishes and one has to deal with coupled, homogeneous, linear PDE's for $\chi_k^{(2)}, \chi_k^{(2)*}$.

Consider the homogeneous system first, since this already provides us with valuable information about the possible outcome of a baryon-baryon collision
in the massive GN model. We discuss separately the situation before and after the collision.
The incoming channel consists of two well separated DHN baryons. Here, we already know that there are no bare mass corrections, except for the relationship
between $y$ and $\nu$. Hence we can assume the initial condition $\chi_k^{(2)} = 0 $ for $t\to - \infty$ for $k=1,2$. After the collision, the possible final states
are determined by the non-trivial solutions of the homogeneous system
\begin{eqnarray}
& & \left(  i \partial_t  + \frac{1}{2} \partial_x^2  + 2y \sum_{\ell} |\psi_{\ell}^{(0)}|^2 \right) \chi_k^{(2)} 
\nonumber \\
&  = &    -2 y \psi_k^{(0)} \sum_{\ell} \left( \psi_{\ell}^{(0)} \chi_{\ell}^{(2)*} + \psi_{\ell}^{(0)*} \chi_{\ell}^{(2)} \right) .
\label{D5}
\end{eqnarray}
Since the TDHF approach cannot give the complete information about individual reaction channels but treats them only in an average way (due to
the assumption of a single Slater determinant), we expect a superposition of different final states. The weights of specific states can only 
be determined by solving the full, inhomogeneous system of PDE's (\ref{D2}) numerically. The homogeneous system (\ref{D5}) can actually be
solved analytically as follows. Suppose we can find a small deformation of the unperturbed solution $\psi_k^{(0)}$ of the NLS equation (\ref{C10})
such that the result is again a solution,
\begin{equation}
\left( i \partial_t + \frac{1}{2} \partial_x^2 + 2 y \sum_{\ell} |\psi_{\ell}^{(0)} + \delta \psi_{\ell}^{(0)}|^2 \right) \left(\psi_k^{(0)} + \delta \psi_k^{(0)} \right) = 0
\label{D6}
\end{equation}  
Linearizing Eq.~(\ref{D6}) in $\delta \psi_k^{(0)}$ yields a solution of Eq.~(\ref{D5}), namely $\chi_k^{(2)} = \delta \psi_k^{(0)}$. 
If the deformation can be taken to be infinitesimal, i.e., if the solution $\psi_k^{(0)}+\delta \psi_k^{(0)}$ is continuously connected to the 
unperturbed solution $\psi_k^{(0)}$, this solution will be exact. Thus, in order to survey the possible asymptotic solutions $\chi_k^{(2)}$,
all we have to do is list the solutions of the NLS equation that can be obtained by a continuous deformation of the unperturbed solution (\ref{D4}).
This should already enable us to characterize the possible final states in a baryon-baryon collision of the massive GN model.

Consider the elastic channel first. A forward scattered baryon will emerge with a time delay and a phase factor. The first change corresponds to
the solutions
\begin{equation}
\chi_k^{(2)} = A_x \partial_x \psi_k^{(0)}, \quad \chi_k^{(2)} = A_t \partial_t \psi_k^{(0)}
\label{D7}
\end{equation} 
with real coefficients $A_{x,t}$. To confirm that these are indeed exact solutions of (\ref{D5}), start from the NLS equation for $\psi_k^{(0)}$ and
just differentiate this equation with respect to $x$ or $t$. In order to describe the phase shift in an analogous way, multiply $\psi_k^{(0)}$ by
$e^{i\delta_k}$ and insert it into the NLS equation. Differentiating with respect to $\delta_k$ then yields another solution of (\ref{D5}),
\begin{equation}
\chi_k^{(2)} = i A_k  \psi_k^{(0)},
\label{D8}
\end{equation}
again with real $A_k$. Clearly, the solutions (\ref{D7},\ref{D8}) cannot account for elastic backward scattering, expected in the massive
GN model. It is not hard to find the corresponding deformation. The multi-component NLS equation
\begin{equation}
\left( i \partial_t + \frac{1}{2} \partial_x^2 + 2 y \sum_{\ell} |\psi_{\ell}^{(0)} |^2 \right) \psi_k^{(0)} = 0
\label{D9}
\end{equation}
remains valid under unitary transformations of the $\psi_k$, here under the group U(2) since $k=1,2$ only. An infinitesimal U(2) transformation
can be parameterized as
\begin{equation}
\delta \psi_k^{(0)} = i \left( \varphi {\rm \ 1}+ \vec{\theta} \cdot \vec{\tau}\right)_{k \ell} \psi_{\ell}^{(0)}.
\label{D10}
\end{equation}
The U(1) part and the $\tau_3$ rotation have already been accounted for by (\ref{D8}), so that the only new solution we get is
\begin{equation}
\chi_1^{(2)} = C \psi_2^{(0)}, \quad \chi_2^{(2)} = C^* \psi_1^{(0)}
\label{D11}
\end{equation}
with complex coefficient $C$. Since $\psi_{1}^{(0)}$ and $\psi_2^{(0)}$ are moving in opposite directions at the same speed, this is exactly what it takes 
to describe elastic backward scattering.

We now look for deformations of $\psi_k^{(0)}$ related to inelastic processes. The simplest possibility is that the baryon changes its
velocity, i.e.,
\begin{equation}
\chi_k^{(2)} = A_v \partial_v \psi_k^{(0)}
\label{D12}
\end{equation}
with real $A_v$. The baryon may also change its size parameter (and thereby fermion number). Due to the factor of $2y$ in the potential $S^{(2)}$,
we cannot simply differentiate $\psi_k^{(0)}$ with respect to $y$ in this case. A simple calculation shows that the following modified
expression is an exact solution of (\ref{D5}),
\begin{equation}
\chi_k^{(2)} = A_y (2y + \partial_y) \psi_k^{(0)}.
\label{D13}
\end{equation} 
Presumably, this does not yet exhaust all possibilities. It is known that the (multi-component) NLS equation possesses solutions with more than
one soliton in each component \cite{17}. We have not found a simple way of relating these multi-soliton solutions continuously to the standard solution,
but cannot rule out such a possibility. Then these more complicated solutions of the NLS equation might also play some role in inelastic processes.
In the present work, we shall not consider multi-soliton deformations any further, but see how far we can get with the above, simplest solutions. 

Eventually, we must solve the inhomogeneous system of PDE's (\ref{D2}) numerically. Since this problem amounts to solving an
inhomogeneous, time dependent Schr\"odinger equation with time dependent Hamiltonian, a number of numerical
methods are available in the literature. We follow the method described by Puzynin et al. \cite{18}. It is a higher-order stable operator-difference scheme,
generalizing the Crank-Nicolson scheme. It can be derived from the Magnus expansion of the evolution operator for one time step.
We actually used the 2nd order variant described in detail in Ref.~\cite{18}. In order to solve the equation
\begin{equation}
i \frac{d\psi(t)}{dt} = H(t) \psi(t)+ Q(t),
\label{D14}
\end{equation} 
one divides the temporal interval $[0,T]$ into $K$ steps of length $\tau$ ($t_k=k \tau, k=0,1,...,K$). We also need the intermediate
points $t_{k+1/2} = t_k+\tau/2$ and denote $\psi^k = \psi(t_k), \psi^{k+1/2}=\psi(t_{k+1/2})$.
Define
\begin{widetext}
\begin{eqnarray}
F_k & = & H(t_{k+1/2}) + \frac{\tau^2}{24} \ddot{H}(t_{k+1/2})- i \frac{\tau^2}{12} \left[ \dot{H}(t_{k+1/2}), H(t_{k+1/2}) \right],
\nonumber \\
Q_1^k & = & \frac{1}{2} Q(t_k) + \frac{\tau}{12} \left[ \dot{Q}(t_k) + i H(t_k)Q(t_k) \right],
\nonumber \\
Q_2^k & = & \frac{1}{2} Q(t_{k+1}) - \frac{\tau}{12} \left[ \dot{Q}(t_{k+1}) + i H(t_{k+1})Q(t_{k+1}) \right].
\label{D15}
\end{eqnarray}
Then the step from $t_k$ to $t_{k+1}$ goes as follows,
\begin{eqnarray}
\left( 1- \frac{1}{4} \tau \alpha F_k \right) \psi^{k+1/2} & = & \left(1-\frac{1}{4} \tau \alpha^* F_k \right) \left( \psi^k -i \tau Q_1^k \right),
\nonumber \\
\left( 1+ \frac{1}{4} \tau \alpha^* F_k \right) \left( \psi_{k+1}  + i \tau Q_2^k \right)    & = & \left(1+\frac{1}{4} \tau \alpha F_k \right) \psi^{k+1/2} ,
\label{D16}
\end{eqnarray}
\end{widetext}
with $\alpha = 1/\sqrt{3}-i$. If we choose a spatial grid with $M$ points, $H$ and $F$ will be $4M \times 4M$ 
matrices due to the coupling of $\chi_1, \chi_2, \chi_1^*, \chi_2^*$. Similarly, $\psi$ and $Q_{1,2}$ are $4M$ component vectors. Consequently, Eqs.~(\ref{D16}) are two
systems of $4M$ linear, inhomogeneous algebraic equations which can be solved by standard methods. In the particular case at hand, due to the symmetry of the
scatterers and of the initial conditions, it is actually possible to reduce the dimension by a factor of 2 by restricting $x$-space to a half line.
The number of mesh points in time and space were chosen such that a further increase would not show up in the figures below. We found that 
250 points on the time axis and 250 points on half the space axis were adequate. In this case, all numerical computations could still be done with Maple.

\section{Results}
\label{sect5}

\begin{figure}
\begin{center}
\epsfig{file=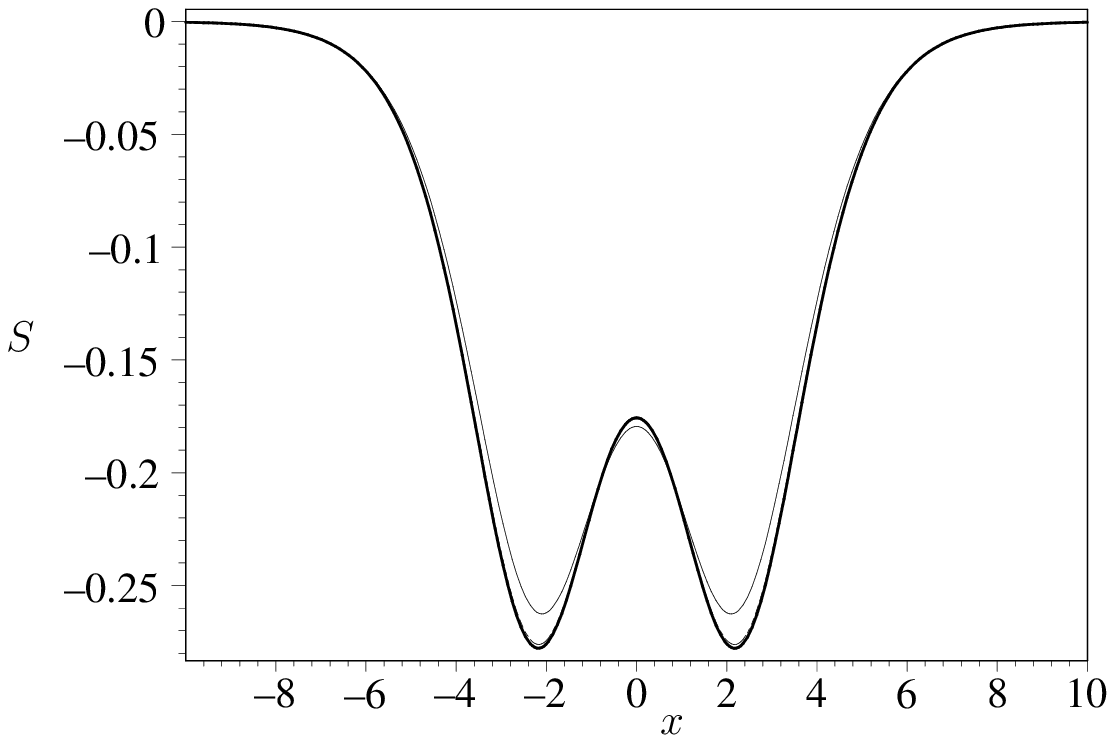,width=8cm,angle=0}
\caption{Scalar potential $S$ for baryon-baryon scattering at $\gamma=0$. Parameters: $y=0.5, v=0.3, t=5.0$.
Thin curve: LO, dashed: LO+NLO, fat: full calculation. All curves have been calculated analytically.}
\label{fig1}
\end{center}
\end{figure}

\begin{figure}
\begin{center}
\epsfig{file=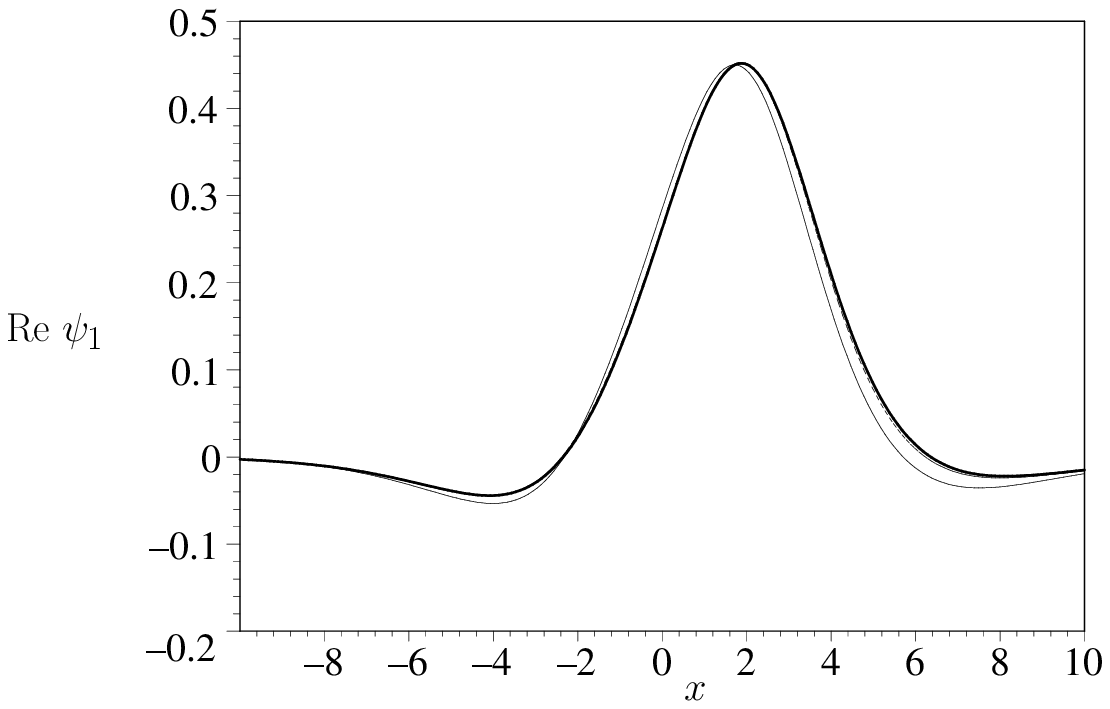,width=8cm,angle=0}
\caption{Like Fig.~\ref{fig1}, but real part of bound state wave function $\psi_1$ (large component).}
\label{fig2}
\end{center}
\end{figure}

\begin{figure}
\begin{center}
\epsfig{file=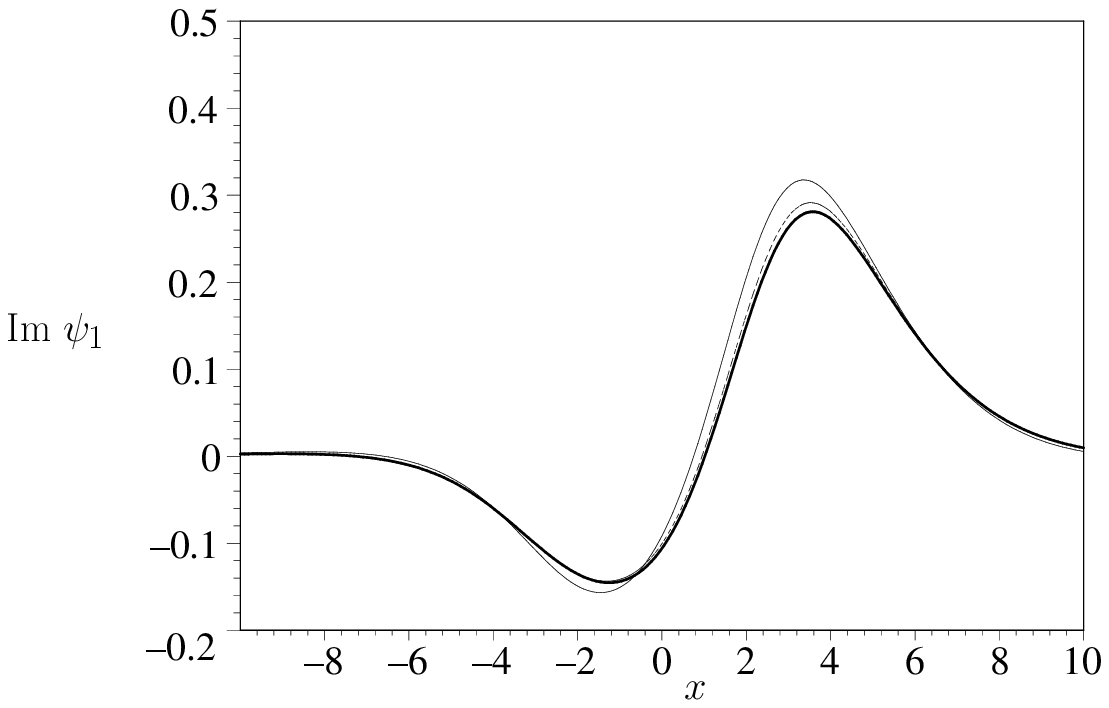,width=8cm,angle=0}
\caption{Like Fig.~\ref{fig1}, but imaginary part of bound state wave function $\psi_1$ (large component).}
\label{fig3}
\end{center}
\end{figure}

A necessary condition for being able to trust the results is that we work in a regime where the non-relativistic expansion
converges well in the massless limit. Since we know the exact result in this case, this is something which can be checked. To this end, we have computed 
the scalar mean field $S$ and the bound state spinors $\psi_{1,2}$ for scattering of two DHN baryons in the cm frame analytically. We then introduce
the parameter $\epsilon$ ($y \to \epsilon y, v \to \epsilon v, x \to x/\epsilon, t \to t/\epsilon^2$) and expand potential and spinors in $\epsilon$ to NLO.
The resulting expressions are too long to be shown here, but can readily be generated by Computer Algebra. We have also checked that this indeed solves 
Eq.~(\ref{C11}) at $\gamma=0$, a useful further test of the above formalism. We then compare the exact results with the LO and NLO approximations
by looking at animations of the corresponding plots throughout the whole collision process. We identify the region in ($y,v$)-parameter space 
where the difference between LO and NLO is of the order of 10$\%$, but the difference between NLO and the exact result of the order of a few $\%$ only.
One example of a snapshot of such a survey is shown in Figs.~\ref{fig1},\ref{fig2},\ref{fig3} for $y=0.5, v=0.3$. The result of this search is the region
$0.4 \le y \le 0.6$, $0.1 \le v \le 0.6$. There we are in a situation where the NLO approximation is quantitatively reliable in the massless limit.
The corrections which we compute for the non-integrable, massive model then also have a good chance of being trustworthy. 
We cover this preferred parameter region with 18 points in steps of $\Delta y=0.1, \Delta v =0.1$. Outside this region, the NLO corrections are either negligible
or too big to truncate the non-relativistic expansion after two terms.   

We now turn to the results of solving the inhomogeneous system of PDE's (\ref{C13}) numerically. In the following subsections, we discuss backward and forward scattering   
separately.

\subsection{Backward scattering}
\label{sect5a}

Backward scattering has the unique feature that the result for $\gamma=0$ is strictly zero, due to the integrability of the massless GN model. 
Hence the correction proportional to $\chi_k^{(2)}$ in Eq.~(\ref{C12}) represents the full wave function $\psi_k$ in this region. 
Since this is of O($\epsilon^2$), the density of backward scattered fermions resulting from the $k$-th bound state,
\begin{equation}
\rho_k = \left( \frac{\gamma}{1+\gamma} \right)^2 |\chi_k^{(2)}|^2,
\label{E1}
\end{equation}
is correct, although it is of O($\epsilon^4$) and we have only been working to O($\epsilon^2$). Integration over the negative half-axis then 
yields the reflection coefficient $R$ and a ``reduced" reflection coefficient $R'$,
\begin{equation}
R = \left( \frac{\gamma}{1+\gamma} \right)^2 R', \quad R' = \int_{-\infty}^0 dx |\chi_k^{(2)}|^2.
\label{E2}
\end{equation} 
These are also of O($\epsilon^4$), but can be trusted for the same reason.

\begin{figure}
\begin{center}
\epsfig{file=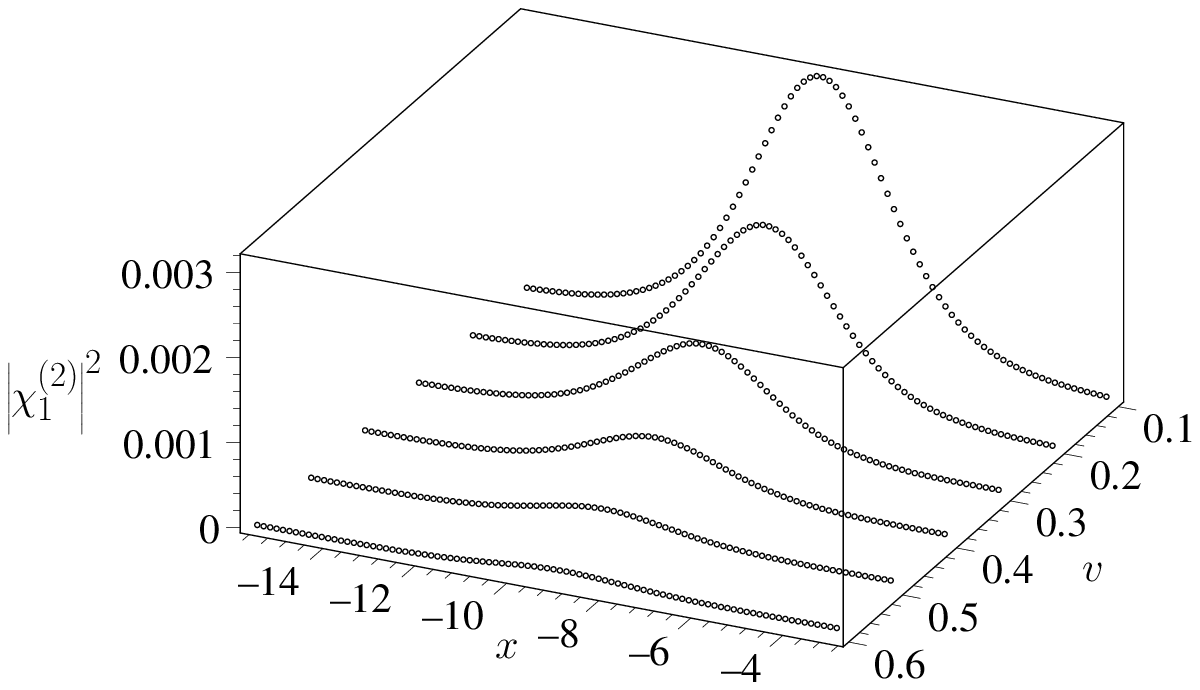,width=8cm,angle=0}
\caption{Reduced density for backscattered fermions after the collision, at $y=0.5$ and different velocities. Numerical results.}
\label{fig4}
\end{center}
\end{figure}

In Fig.~\ref{fig4}, we give an overview of our results for the backscattered, ``reduced" density (i.e., expression (\ref{E1}) without the $\gamma$-dependent factor)
for $y=0.5$, after the collision. At the lower velocities, one sees a clear peak, gradually decreasing towards the highest velocity. 
For $y=0.4$ and $y=0.6$, the results look similar and need not be shown here.

\begin{figure}
\begin{center}
\epsfig{file=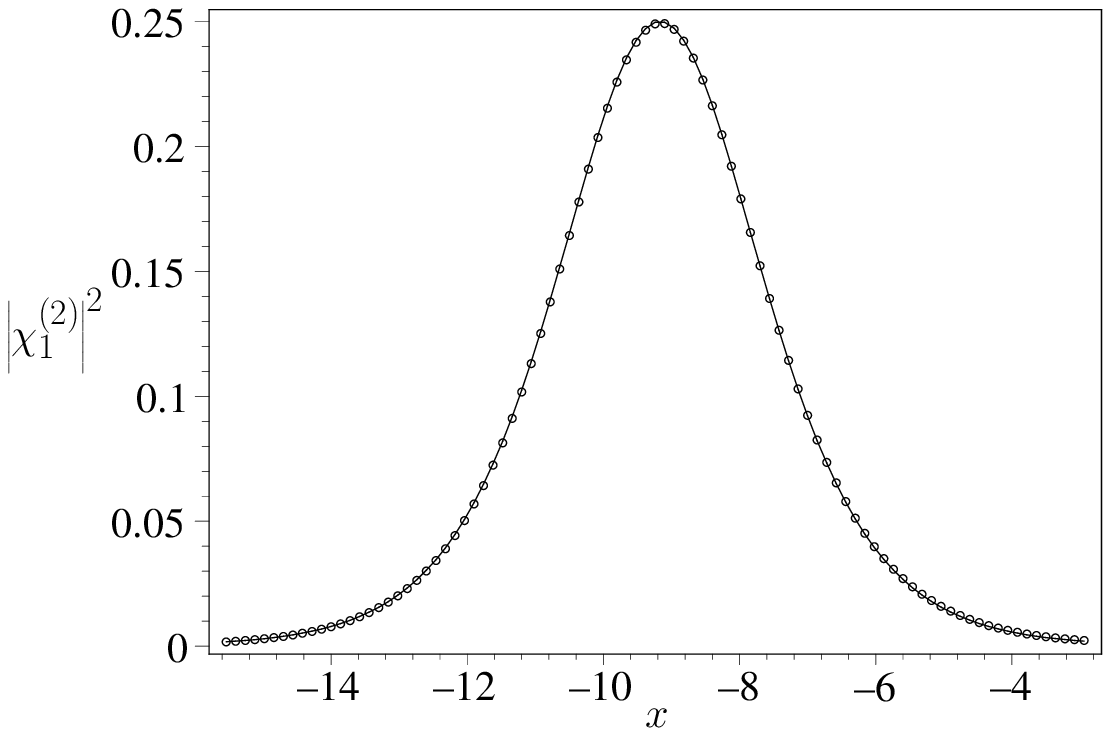,width=8cm,angle=0}
\caption{Fitting the reduced density of backscattered fermions in state 1 at $y=0.5, v=0.1, t=75.4$. Line: LO density of state 2, $|\psi_2^{(0)}|^2$. Points:
numerical calculation, multiplied by a factor of 79.}
\label{fig5}
\end{center}
\end{figure}

\begin{figure}
\begin{center}
\epsfig{file=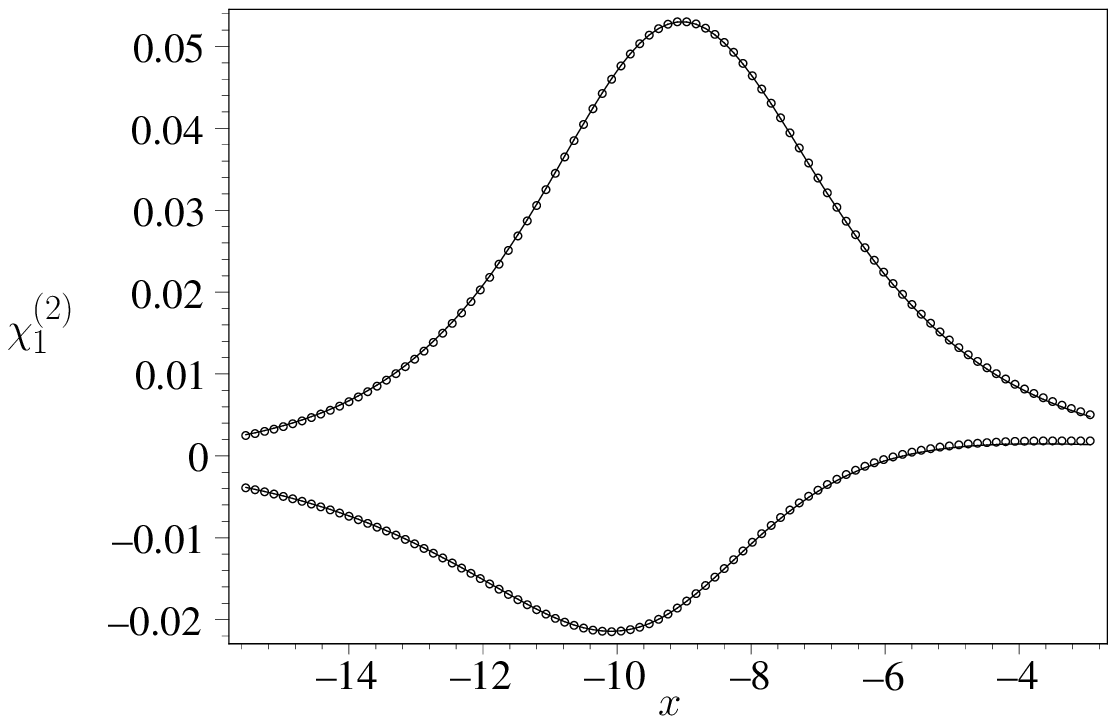,width=8cm,angle=0}
\caption{Like Fig.~\ref{fig5}, but imaginary part (upper dots) and real part (lower dots) of $\chi_1^{(2)}$ shown. Solid lines: Real and imaginary parts of $C \psi_2^{(0)}$
with best fit parameter $C=0.11-0.022i$.}
\label{fig6}
\end{center}
\end{figure}

In order to understand the nature of the peak, we confront it with our theoretical expectation based on solutions of the homogeneous PDE's (\ref{D5}). 
The only candidate for backward scattering which we have identified in Sect.~\ref{sect4} is Eq.~(\ref{D11}) describing elastic scattering. In Figs.~\ref{fig5} 
and \ref{fig6} we compare the numerically calculated $|\chi_1^{(2)}|^2$ and $\chi_1^{(2)}$ at $y=0.5$ and the lowest velocity, $v=0.1$, to (\ref{D11}). 
We get perfect agreement between numerical calculation and analytical prediction for the value $C= 0.110-0.022 i$.
The fact that backward scattering at the lowest energy is purely elastic is confirmed by comparable
fits at the other two values of $y$, yielding $C=0.066-0.016i$ at $y=0.4$ and $C=0.167-0.027i$ at $y=0.6$.

\begin{figure}
\begin{center}
\epsfig{file=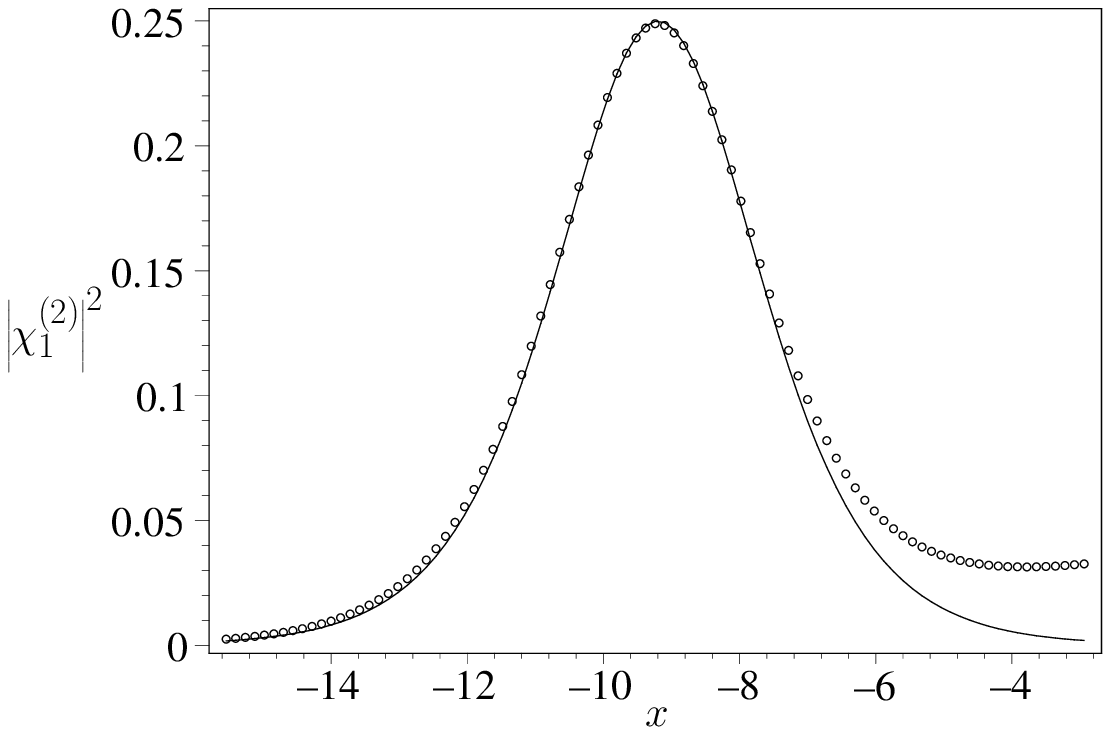,width=8cm,angle=0}
\caption{Like Fig.~\ref{fig5}, but at higher velocity ($y=0.5,v=0.4,t=21.8$). Line: $|\psi_2^{(0)}|^2$. Points:
numerical calculation, multiplied by a factor of 439.}
\label{fig7}
\end{center}
\end{figure}

\begin{figure}
\begin{center}
\epsfig{file=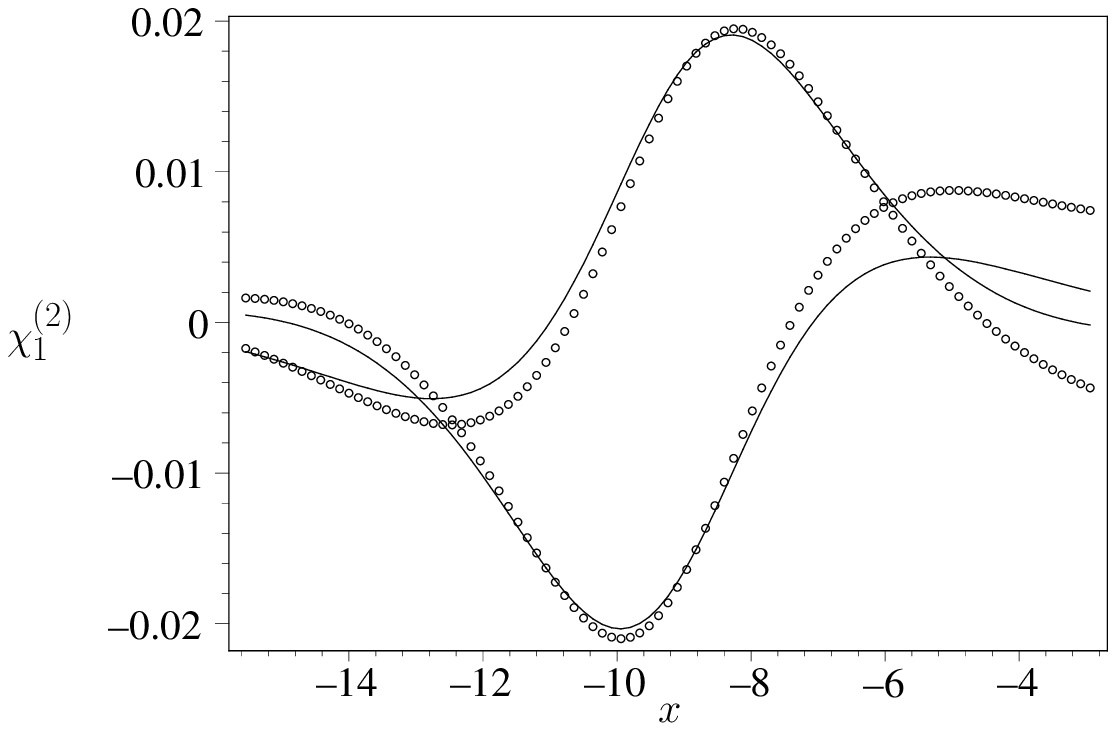,width=8cm,angle=0}
\caption{Same parameters as Fig~\ref{fig7}, but imaginary (upper points) and real part (lower points) of $\chi_1^{(2)}$. Solid lines: Real and imaginary parts of $C \psi_2^{(0)}$
with best fit parameter $C=0.037-0.030i$.}
\label{fig8}
\end{center}
\end{figure}

\begin{figure}
\begin{center}
\epsfig{file=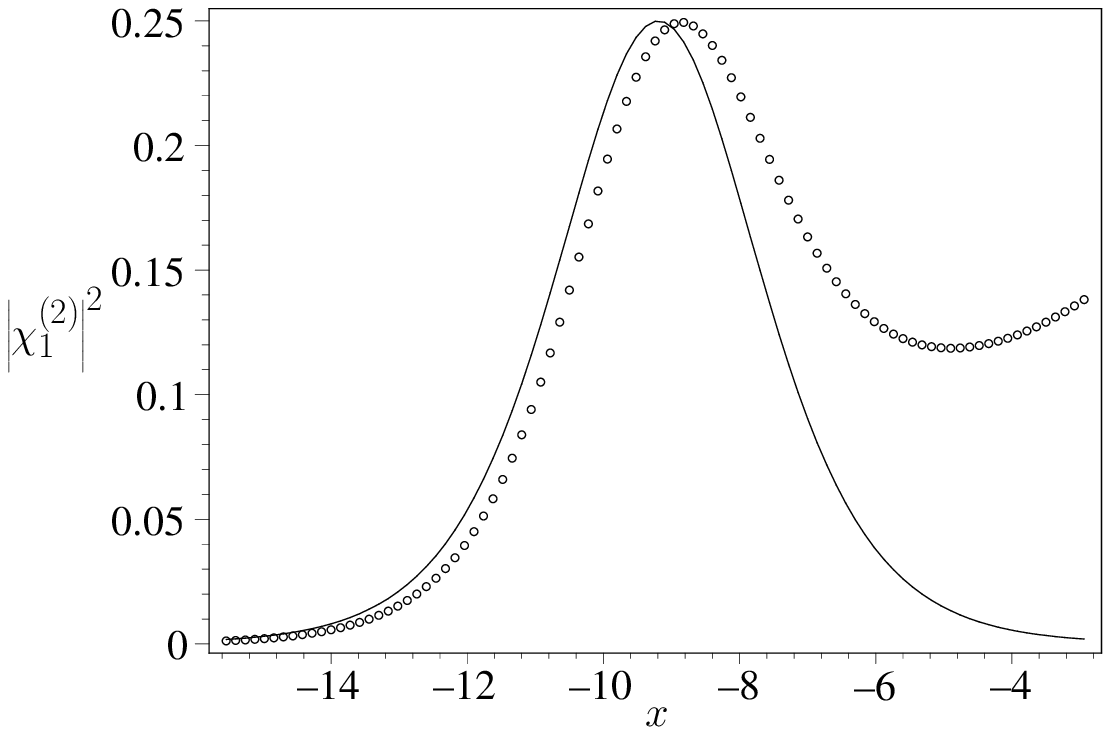,width=8cm,angle=0}
\caption{Like Fig.~\ref{fig7}, but at the highest velocity considered ($y=0.5,v=0.6,t=14.9$) to illustrate rise of inelastic background. Line: $|\psi_2^{(0)}|^2$. Points:
numerical calculation, multiplied by a factor of 1780.}
\label{fig9}
\end{center}
\end{figure}

\begin{figure}
\begin{center}
\epsfig{file=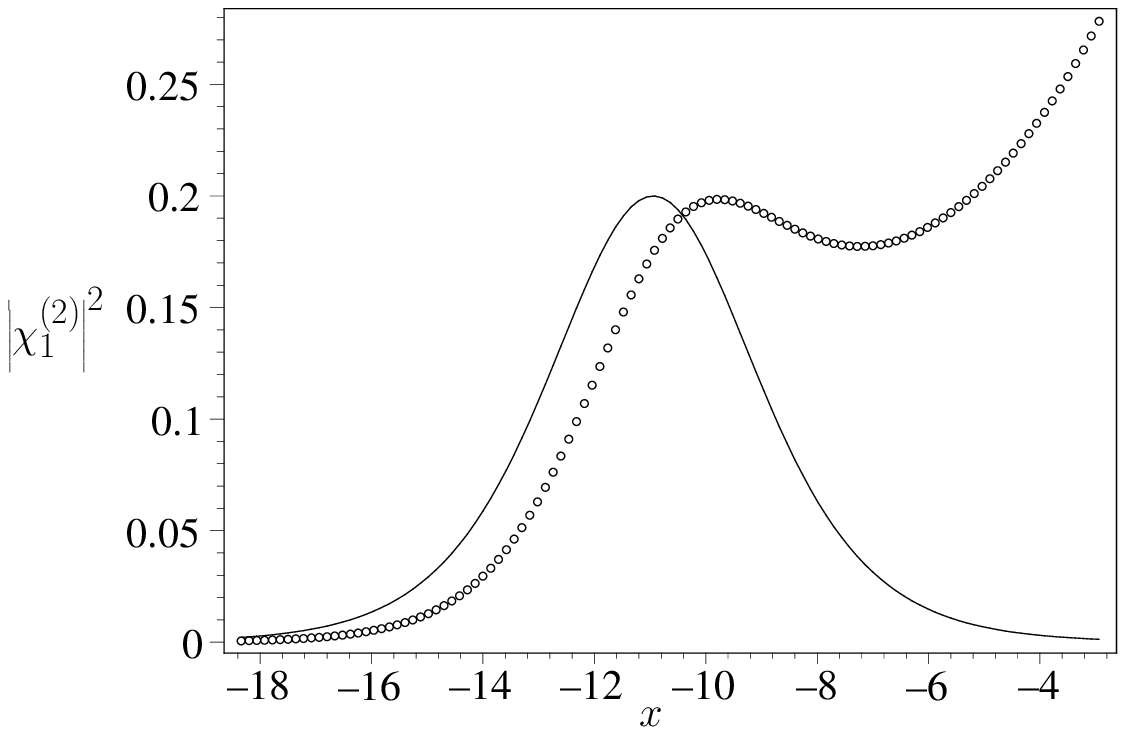,width=8cm,angle=0}
\caption{Like Fig.~\ref{fig9}, but at lowest $y$ value ($y=0.4,v=0.6,t=17.9$). Line: $|\psi_2^{(0)}|^2$. Points:
numerical calculation, multiplied by a factor of 12500.}
\label{fig10}
\end{center}
\end{figure}

\begin{figure}
\begin{center}
\epsfig{file=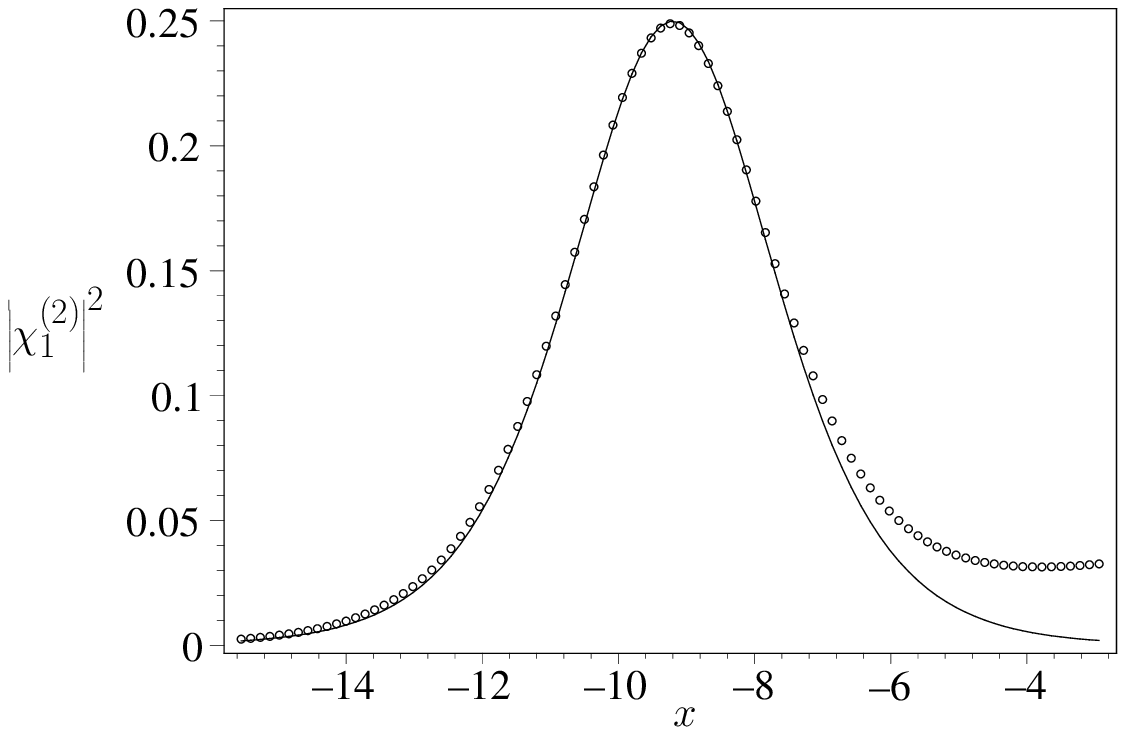,width=8cm,angle=0}
\caption{Like Fig.~\ref{fig9}, but at highest $y$ value ($y=0.6,v=0.6,t=12.9$). Line: $|\psi_2^{(0)}|^2$. Points:
numerical calculation, multiplied by a factor of 415.}
\label{fig11}
\end{center}
\end{figure}

If we move to higher velocities, a background is appearing below the elastic peak, see Fig.~\ref{fig7} for an example of the reduced density at $y=0.5, v=0.4$. At the same time,
the fit to the elastically backscattered wave function deteriorates, see Fig.~\ref{fig8}. The background can only be due to the onset of inelastic processes
at higher energies, but so far we are lacking a way of analyzing it quantitatively. At $y=0.5$ and the highest velocity, $v=0.6$, the background becomes even more prominent, see Fig.~\ref{fig9},
and a fit with the elastic wave function impossible due to interference with the large, unknown background. To illustrate the $y$ dependence, we 
show the corresponding plots at $y=0.4$, Fig.~\ref{fig10}, and $y=0.6$, Fig.~\ref{fig11}. Evidently, the ratio of elastic peak to inelastic background increases with decreasing $y$.
This is what one would expect qualitatively, since the fermions are more loosely bound for smaller $y$, cf. Eq.~(\ref{C15}). 

\begin{figure}
\begin{center}
\epsfig{file=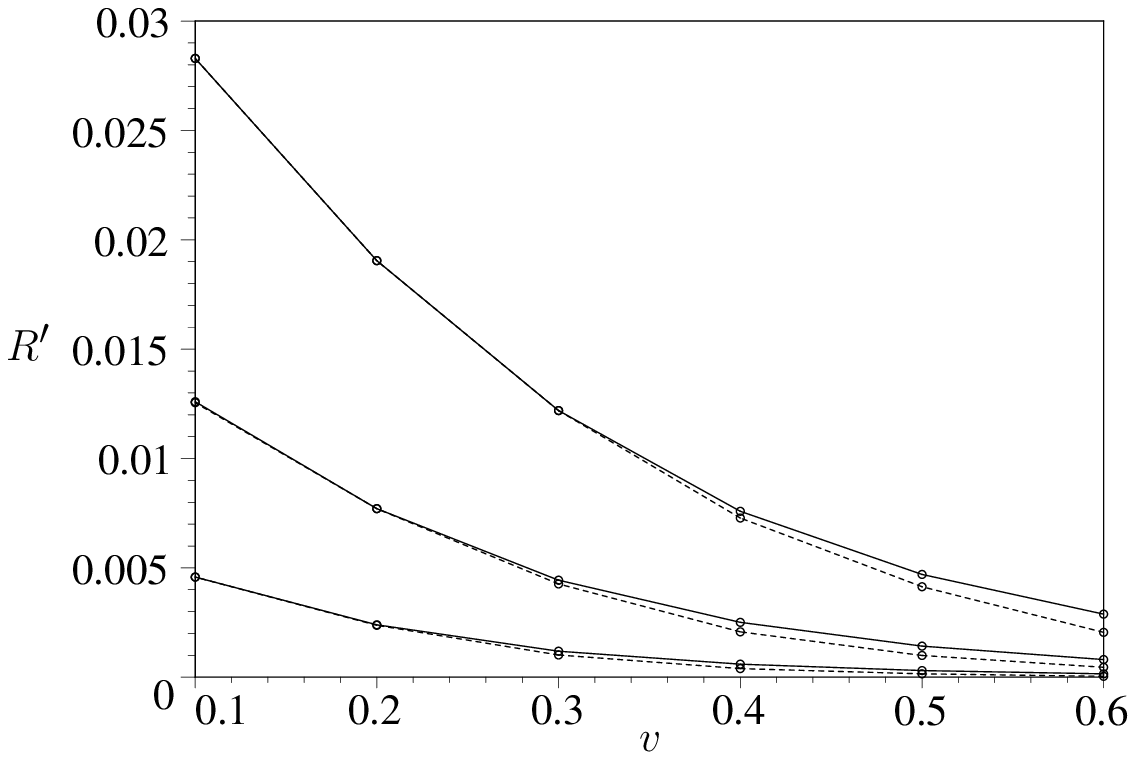,width=8cm,angle=0}
\caption{Reduced reflection coefficient $R'$, related to reflection coefficient $R$ by a $\gamma$-dependent factor, see Eq.~(\ref{E2}). From top to bottom: $y=0.6,0.5,0.4$.
Upper (solid) curves: Including background, lower (dashed) curves: elastic only. Points are calculated, lines drawn to guide the eye.}
\label{fig12}
\end{center}
\end{figure}

Finally, Fig.~\ref{fig12} shows the reduced reflection coefficient $R'$ for all calculated points. It differs from the reflection coefficient $R$ by a 
$\gamma$-dependent factor, see Eq.~(\ref{E2}). It is impossible to neatly separate elastic scattering from the inelastic background, since everything is coherent
in TDHF. To get at least a rough idea, we have fitted the smooth background by a 2nd order polynomial in $x$  and subtracted it, assuming incoherence.
Fig.~\ref{fig12} shows both the total and elastic values of $R'$ thus obtained.

\subsection{Forward scattering}
\label{sect5b}

\begin{figure}
\begin{center}
\epsfig{file=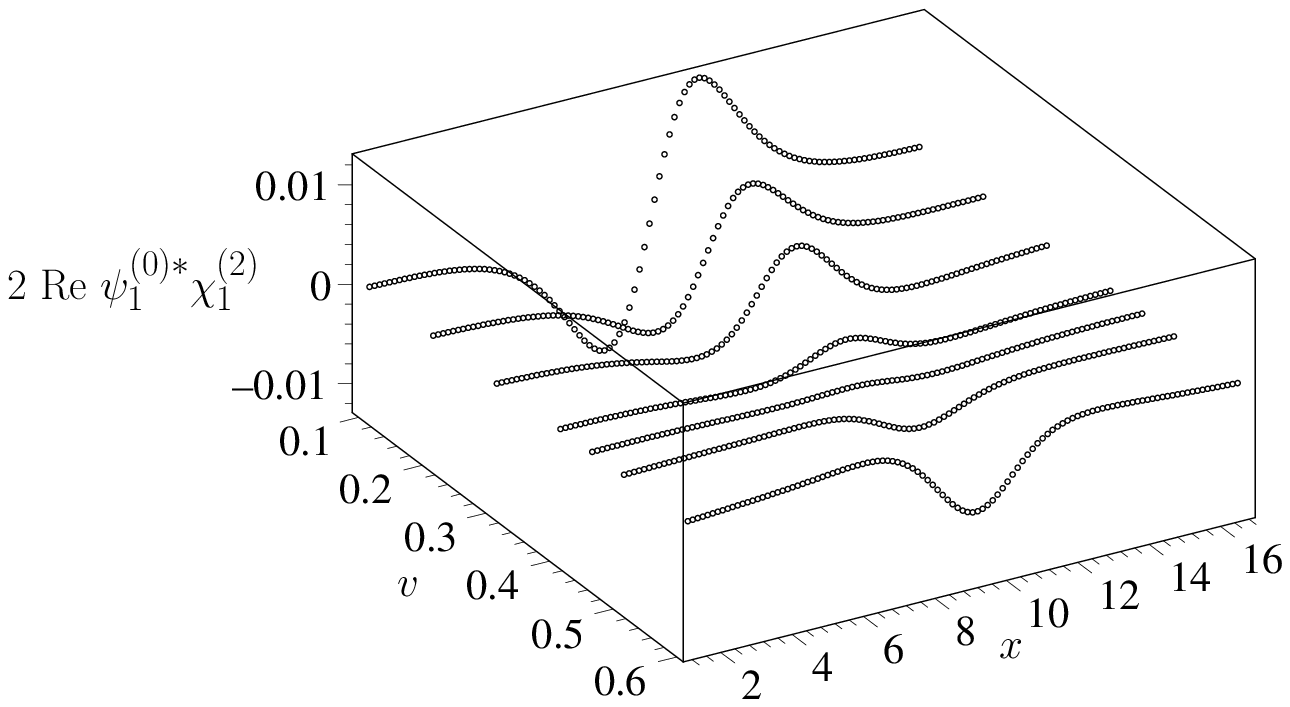,width=8cm,angle=0}
\caption{Reduced change in fermion density due to bare mass, Eq.~(\ref{E3}), forward scattering. The points are computed numerically at $y=0.5$ and
all velocities considered.}
\label{fig13}
\end{center}
\end{figure}

In the case of forward scattering it would not make sense to plot the density $|\chi_k^{(2)}|^2$. Since the LO contribution is non-vanishing, a quantity 
of O($\epsilon^4$) cannot be trusted here. The relevant physical observable of O($\epsilon^2$) is the NLO change in density due to the bare fermion mass,
\begin{equation}
\delta \rho_1  =  \left( \frac{\gamma}{1+\gamma} \right)^2 \delta \rho_1', \quad \delta \rho_1' =  \psi_1^{(0)} \chi_1^{(2)*} + \psi_1^{(0)*} \chi_1^{(2)}. 
\label{E3}
\end{equation}
We first give an overview of our numerical results for $\delta \rho_1'$ at $y=0.5$ in Fig.~\ref{fig13}. 
The solution of the TDHF equation yields the whole time evolution of $\delta \rho_1'$, but here we only show
a snapshot taken after the collision.  As $v$ increases, the curves transform from an   
antisymmetric into a symmetric shape. In order to highlight the transition between these two extrema, we have included one extra point along the $v$-axis
where $\delta \rho_1'$ is close to zero ($v=0.45$). The shapes of the extremal curves 
at $v=0.1$ and $v=0.6$ have a simple interpretation. Since $\delta \rho_1'$ is the difference between two bell-shaped densities, at $v=0.1$ the curves indicate 
a spatial shift between the densities, as expected in purely elastic scattering. The shape at $v=0.6$ on the other hand is suggestive of a
change in the width of the density, consistent with inelastic reactions where the size parameter $y$ (and therefore fermion number) changes.

\begin{figure}
\begin{center}
\epsfig{file=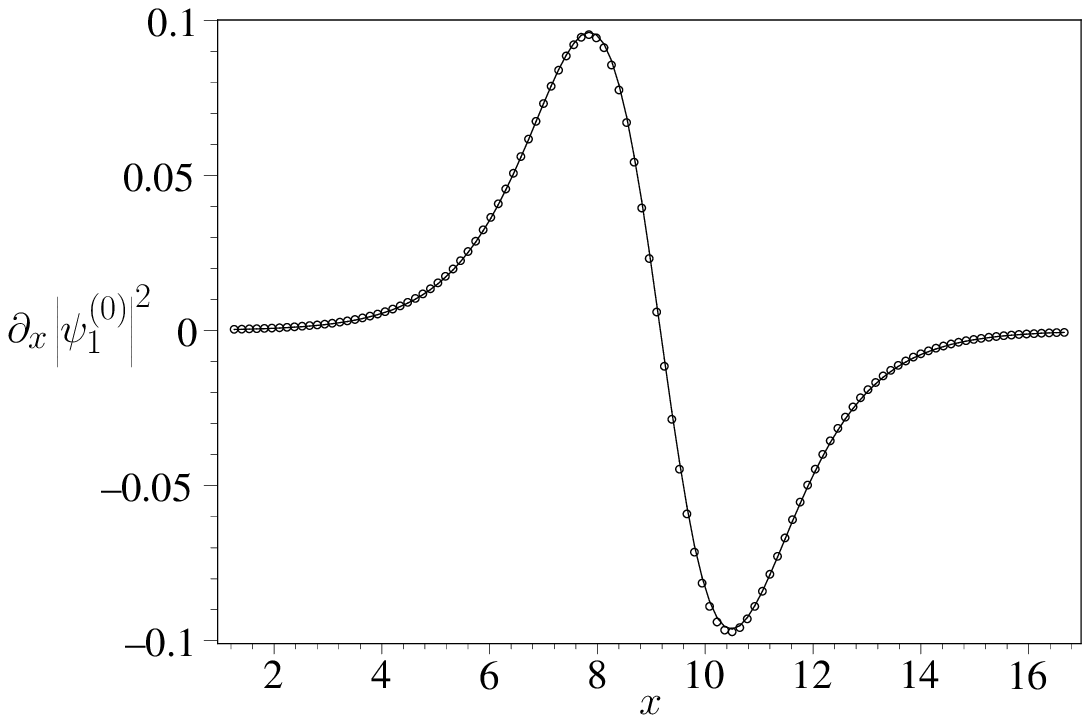,width=8cm,angle=0}
\caption{Fitting the reduced change of density for forward scattering with ansatz implying purely elastic scattering. Parameters: $y=0.5,v=0.1,t=75.4$.
Line: $\partial_x |\psi_1^{(0)}|^2$, points: numerical results for $\psi_1^{(0)}\chi_1^{(2)*}+ \psi_1^{(0)*}\chi_1^{(2)}$, rescaled by a factor of $-7.70$.}
\label{fig14}
\end{center}
\end{figure}

\begin{figure}
\begin{center}
\epsfig{file=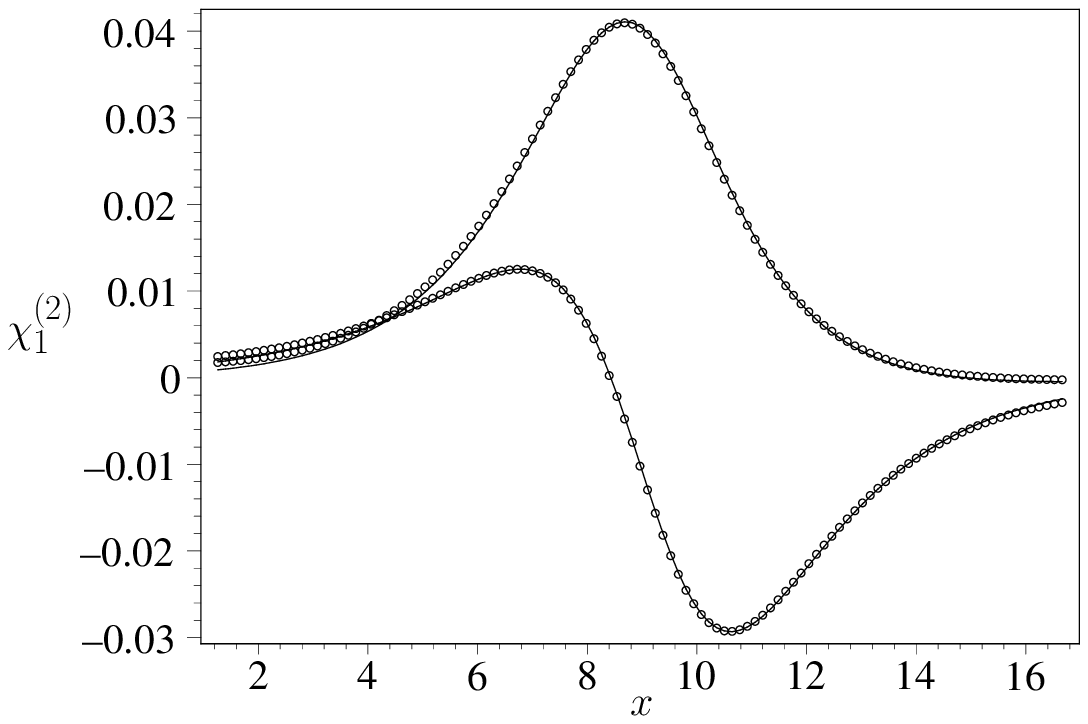,width=8cm,angle=0}
\caption{Like FIG.~\ref{fig14}, but imaginary part (upper points) and real part (lower points) of $\chi_1^{(2)}$. Solid lines: Real and imaginary parts of best fit 
with function $(-0.130 \partial_x - i 0.0707i ) \psi_1^{(0)}$.}
\label{fig15}
\end{center}
\end{figure}

Let us test this interpretation against the analytical, asymptotic predictions discussed in Sect.~\ref{sect4}. If baryon-baryon scattering in the massive GN model
is purely elastic, $\chi_1^{(2)}$ must be proportional to a linear combination of $\partial_x \psi_k^{(0)}$, $\partial_t \psi_k^{(0)}$ and $i \psi_k^{(0)}$ with real
coefficients, see Eqs.~(\ref{D7},\ref{D8}). The first two solutions account for a change in time delay and cannot be distinguished here,
the 3rd accounts for a change in the forward scattering phase shift. Figs.~\ref{fig14} and \ref{fig15} show a corresponding fit at $y=0.5,v=0.1$,
\begin{equation}
\delta \rho' = A \partial_x |\psi_1^{(0)}|^2, \quad \chi_1^{(2)} = ( A \partial_x  + i B) \psi_1^{(0)},
\label{E4}
\end{equation}
with $A=-0.130, B=-0.0707$. Thus at the lowest velocity $v=0.1$, everything is consistent with purely elastic scattering, both in forward and backward direction.

\begin{figure}
\begin{center}
\epsfig{file=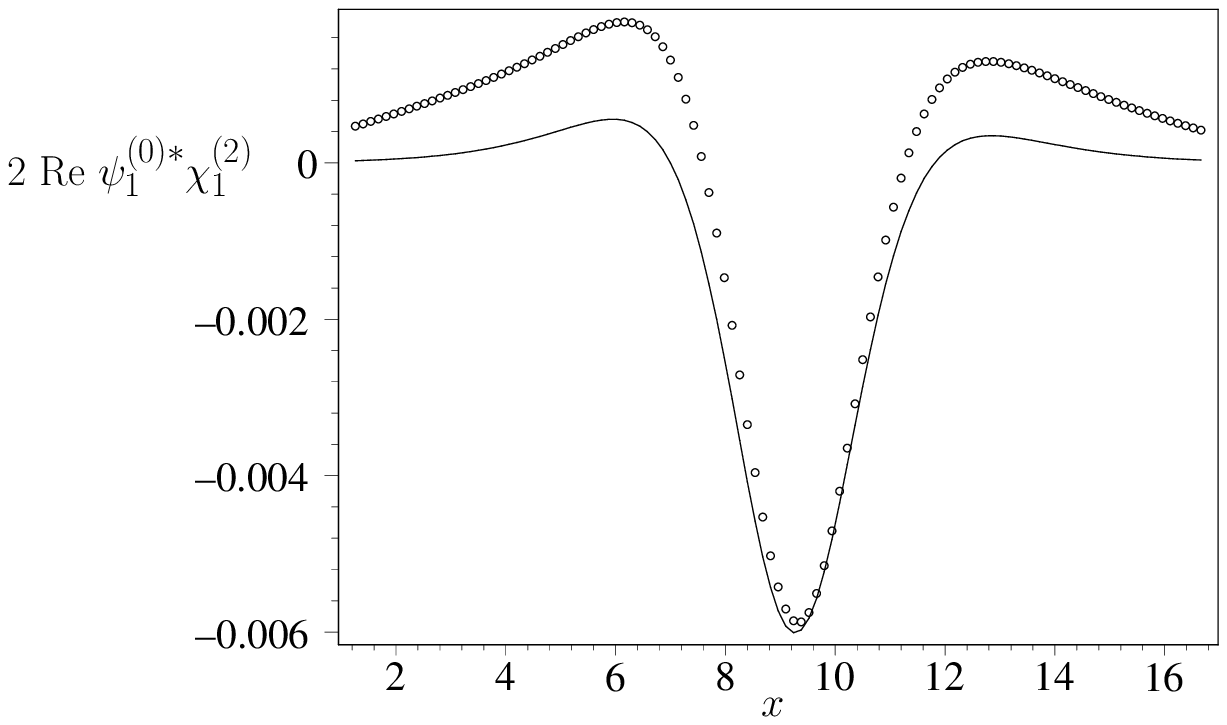,width=8cm,angle=0}
\caption{Attempt to fit the reduced change of density for forward scattering with ansatz implying elastic and inelastic scattering. Parameters: $y=0.5,v=0.6,t=14.9$.
Line: $[0.002 \partial_x +0.006 (4y+\partial_y)] |\psi_1^{(0)}|^2$, points: numerical results for $\psi_1^{(0)}\chi_1^{(2)*}+ \psi_1^{(0)*}\chi_1^{(2)}$.}
\label{fig16}
\end{center}
\end{figure}

At higher values of $v$, it is difficult to find any satisfactory fit with the available basis functions. By way of example, consider $y=0.5, v=0.6$ where inelastic processes
are expected to be important. In Fig.~\ref{fig16}, we have tried to fit $\delta \rho_1'$ with a linear combination of solutions (\ref{D7}) and (\ref{D13}), i.e., 
\begin{equation}
\delta \rho_1' =  \left[ A  \partial_x + B (4y+ \partial_y) \right] |\psi_1^{(0)}|^2.
\label{E5}
\end{equation}
Even the best fit values $A=0.00195, B=-0.00599$ are not satisfactory, although the shape of the curve is qualitatively reproduced. Taking into account the further 
inelastic solution (\ref{D12}) does not improve matters, so that our parametrization of the inelastic solutions is obviously incomplete. As mentioned in
Sect.~\ref{sect4}, one could perhaps generate further candidates by considering multi-soliton solutions of the single component NLS equation, but this is 
left for future work.

So far, all results for forward scattering shown refer to $y=0.5$. We have performed the same kind of calculations at $y=0.4$ and $0.6$ as well. The plots analogous to Fig.~\ref{fig13}
look similar, except that the transition from ``shift" to ``broadening" type shapes happens around $v=0.35$ at $y=0.4$ and $v=0.55$ at $y=0.6$. Thus, elastic
scattering prevails up to higher energies for larger values of $y$, as already noticed in backward scattering. For all 3 values of $y$, the fit (\ref{E4})
is excellent at $v=0.1$, with parameters $A=-0.0826, B=-0.0527$ at $y=0.4$ and $A=-0.150, B=-0.081$ at $y=0.6$, confirming elastic scattering dominance at 
the lowest velocity considered. 

\begin{figure}
\begin{center}
\epsfig{file=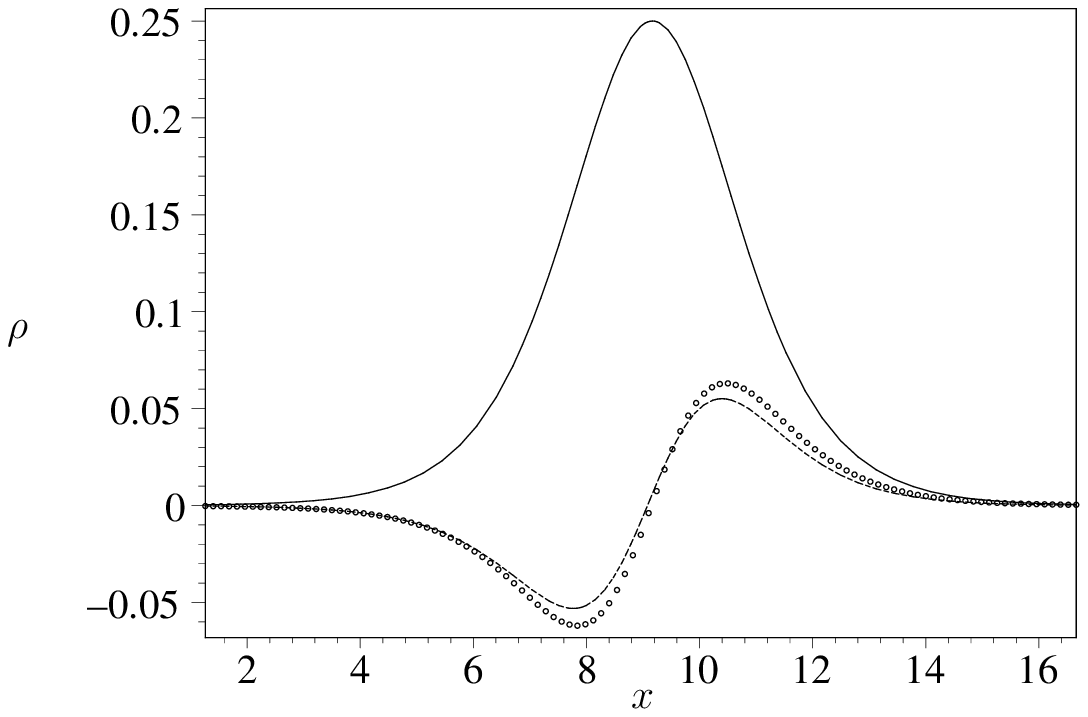,width=8cm,angle=0}
\caption{Solid line: LO fermion density, $|\psi_1^{(0)}|^2$, for forward scattering ($y=0.5, v=0.1, t=75.4$). The other two curves are NLO contributions,
multiplied by a factor of 5 for better visibility. Dashed curve: relativistic corrections at $\gamma=0$, points: bare mass corrections, $\delta \rho_1$
of Eq.~(\ref{E3}).}
\label{fig17}
\end{center}
\end{figure}

\begin{figure}
\begin{center}
\epsfig{file=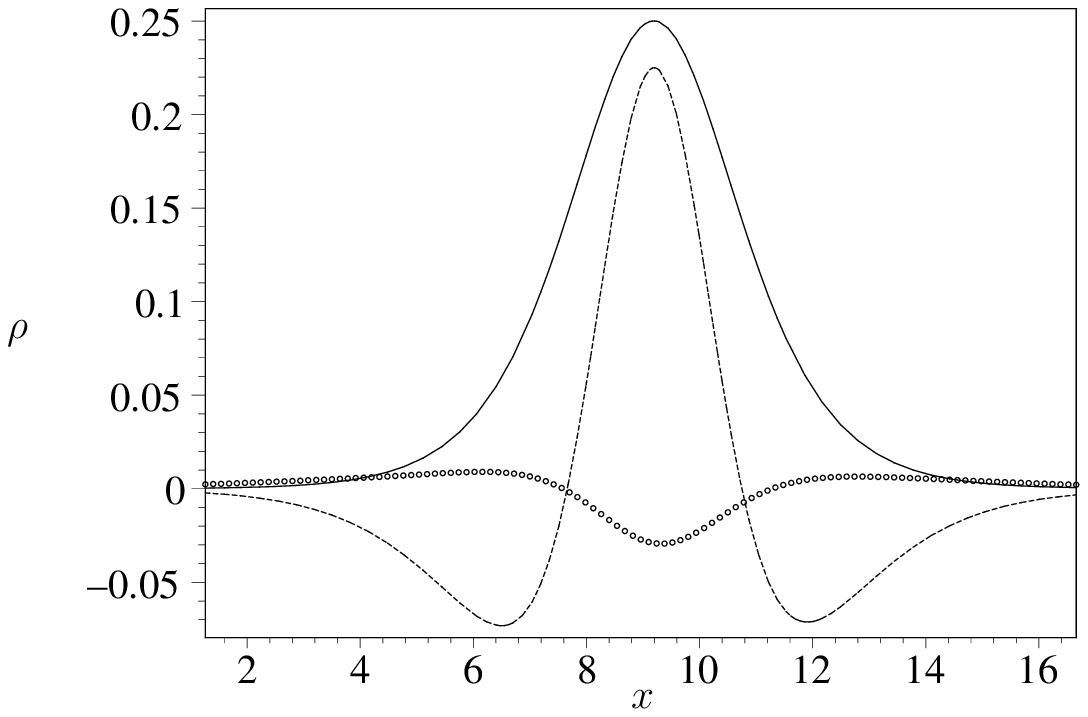,width=8cm,angle=0}
\caption{Like Fig.~\ref{fig17}, but at the highest velocity ($y=0.5,v=0.6,T=14.9$).}
\label{fig18}
\end{center}
\end{figure}

Recall that there are two independent corrections to the fermion density of O($\epsilon^2$). The first one arises from relativistic corrections to the NLS equation 
at $\gamma=0$ (``fine structure" and vacuum polarization type effects), discussed above in the context of Figs.~\ref{fig1},\ref{fig2},\ref{fig3}.
It is present in the massless GN model, does not destroy integrability and can be computed in closed analytical form. The second one is $\delta \rho_1$ 
from Eq.~(\ref{E3}) originating from the bare mass term and to be computed numerically.
It is interesting to compare these two corrections to the LO density and to each other. This is done in Fig.~\ref{fig17} for $y=0.5, v=0.1$ (purely elastic scattering)
and in Fig.~\ref{fig18} for $y=0.5, v=0.6$ (important inelastic contribution). For better visibility, the two NLO corrections have been multiplied by 5. 
In the elastic scattering case, Fig.~\ref{fig17}, the two corrections almost coincide, but this is most likely accidental. Both corrections go into the same direction 
of decreasing the time delay. In Fig.~\ref{fig18} where inelastic scattering plays a role, the relativistic correction is an order of magnitude larger than
the mass correction and has opposite sign, but the shapes of the curves are again similar. The sign of the mass correction is such that the unperturbed density
gets broadened, as one would expect from inelastic reactions where $y$ (and therefore fermion number) should decrease.
 
We finish with a remark about conservation of fermion number. The integral over the $x$-axis of both NLO contributions to the density in forward scattering vanishes, as we
have checked numerically. The LO density is normalized to 1. Hence the total number of forward scattered fermions is the same in the massless and massive models.
Nevertheless, we have computed a reflection coefficient of O($\epsilon^4$) in the massive GN model, whereas it vanishes in the massless limit. There 
is no contradiction between these findings. In forward direction, all we have shown is the vanishing of the O($\epsilon^2$) contribution. In order to check
the expected reduction of the forward scattered fermions of O($\epsilon^4$), it would be necessary to do a full NNLO calculation. This is clearly impossible
with the formalism developed so far.

\section{Summary and conclusions}
\label{sect6}
 
Recent progress on baryon-baryon scattering in the massless GN model
owes much to the fact that this is an integrable quantum field theory. This enables us to compute analytically processes as complex as the collision
between relativistic, composite particles. Unfortunately, the physical phenomena
which can be studied in this manner are also severely limited by integrability. As is often the case, the same facts which make a system exactly solvable 
also render it somewhat unphysical. Only elastic forward scattering is allowed, no matter at what energy one collides which types of projectiles. 
This motivates us to turn our attention to non-integrable toy models promising a richer dynamics, like backscattering and inelastic reactions.
Unlike real particle production, break-up processes should not be suppressed at large $N$, so that the massive GN model is a good candidate 
for exploring the transition from integrable to non-integrable systems with semi-classical methods.
 
Since a full numerical TDHF calculation seems to be exceedingly hard even in such a toy model, we had to compromise and focus 
on the vicinity of the non-relativistic regime. In this case, one can take advantage of the benefits offered by an effective no-sea theory 
applicable here. This enables us to formulate the TDHF problem in a manner close to non-relativistic many-body theory, based on a
time dependent Schr\"odinger equation and systematically calculable relativistic and quantum field theoretic corrections. 
The result of such a calculation is a quantitative
field theoretic calculation with controlled approximations and without adjustable parameters. This paper is actually the first successful
application of the effective no-sea theory of Ref.~\cite{12} to a problem which has not been addressed by any other method before.

Our first important result was rather easy to get at, but is nonetheless quite interesting: In the non-relativistic limit, the multi-component NLS equation
can be used to solve scattering problems both in the massless and massive GN models. To LO, integrability is maintained, even in time
dependent problems. This extends the region of applications where the massive and massless GN models are equally tractable by analytical
methods from static and thermodynamic problems to dynamical problems, at least at very low energies.

The second and more difficult part of the investigation concerns the NLO corrections. This is more interesting, because it shows the onset of
physical processes absent in integrable models, like backscattering and inelastic reactions. Relativistic and quantum field theoretic corrections usually associated with
spectroscopy (fine structure, Lamb shift) now break integrability and induce these forbidden processes. Thanks to our restriction to a certain parameter
range, we could reduce the task to a system of coupled, inhomogeneous, linear PDE's amenable to numerical solution by standard 
methods. The possible final states of a baryon-baryon collision can be obtained analytically by
solving the corresponding homogeneous system. They are related to a certain class of solutions of the multi-channel NLS equation. 

The picture emerging from backward scattering is perhaps the cleanest. We find a prominent elastic peak at low velocities above a smooth
background. This background is steadily rising with increasing energy. Quantitative agreement with density and wave functions at the lowest
velocity has been achieved by assuming purely elastic backward scattering. It is possible to determine the reflection coefficient and
the phase of the backscattered wave function quantitatively. At higher velocities, a similar analysis is hampered by 
our inability to parametrize the inelastic background. A well known problem characteristic of the TDHF approach is the fact that different
reaction channels are hard to disentangle, since they enter in a coherent, average way due to the assumption of a single Slater determinant. 
Interesting findings are the fact that the $\gamma$ dependence is somehow trivial, so that one does not have to repeat the 
calculation for different $\gamma$'s, and that a reflection coefficient of O($\epsilon^4$) can be computed reliably, even though the whole
calculational scheme is truncated at O($\epsilon^2$), This is unique for a correction to an integrable model where 
backscattering vanishes to LO.

In forward scattering, since the LO term does not vanish, the NLO terms are always interference terms and harder to interpret. Again, the cleanest result
is elastic scattering which exhausts what we see at $v=0.1$. We can compute the change in time delay and in scattering phase shift due to
the bare mass, again with a factorized $\gamma$-dependence. The interference with inelastic processes on the other hand is impossible to analyze in detail
with our methods.
At the highest velocity studied, we see qualitatively that the density is broadened in $x$-space, corresponding to lower $y$ or loss of fermion
number. The qualitative change in the density from shift to broadening with increasing energy and at all $y$-values is conspicuous. Thus there is no doubt that we have seen both
backscattering and inelastic reactions in the NLO calculation, unlike at LO.


\end{document}